\documentclass[a4paper,12pt]{article}

\usepackage{jheppub} 

\usepackage[T1]{fontenc} 

\title{\boldmath Bethe/Gauge Correspondence for linear quiver theories with ABCD gauge symmetry and spin chains}

\author[]{Xiang-Mao Ding}
\author[1]{and Tinglyer Zhang \note{Corresponding author.}}

\affiliation[]{Institute of Applied Mathematics, Academy of Mathematics and Systems Science, \\
Chinese Academy of Sciences, Beijing 100190, China}

\emailAdd{xmding@amss.ac.cn}
\emailAdd{zhangting@amss.ac.cn}

\abstract{This note is an extension of \cite{DZ23} there the supersymmetric vacua of three-dimensional $\mathcal{N}=2$ gauge theories with matter are shown to be in one-to-one correspondence with the eigenstate of $\text{XXZ}$ integrable spin chain Hamiltonians with open boundary conditions. We consider the $A_{2}$ quiver gauge theory, which is the simplest non-trivial quiver gauge theory, and $sl_{3}$ open $\text{XXZ}$ spin chain with diagonal boundary condition. We demonstrate the correspondence between the vacuum equations of different gauge groups and Bethe Ansatz equations with different boundary parameters. Not only that, but we furthermore push forward the program to the general $A_{r}$ quiver gauge theory.}

\keywords{Quiver gauge theories, Bethe Ansatz, Supersymmetric}

\begin{document} 
\maketitle
\flushbottom

\section{Introduction}
\label{sec:intro}

The Bethe/Gauge correspondence between the supersymmetry gauge theories and quantum integrable system is a topic of research spanning over a decade \cite{NS09a, MNS00, GS08a, GS08b, NW10, NRS11} and even longer in the background of topological gauge theories \cite{Wit89, GN95, GN94a, GN94b}. The space of supersymmetric vacua is naturally regarded as the state space of a quantum integrable system, whose Hamiltonian is the generator of the (twisted) chiral ring. In other words, the spectrum of the quantum Hamiltonians coincides with the spectrum of the (twisted) chiral ring. Without refer to a specific supersymmetric system, we can say that, for the generator of the (twisted) chiral ring $\mathcal{Q}$, $\mathcal{Q}^{\dagger}$, and Hamiltonian $H$ of the spin chain, they have the relationship $H\sim \mathcal{Q}\mathcal{Q}^{\dagger}+\mathcal{Q}^{\dagger}\mathcal{Q}$. The correspondence between $U(N)$ gauge theories and closed XXX spin chains has been carried out exactly in \cite{NS09a, NS09b}. The vacuum equations of 2d, 3d and 4d $A$-type gauge theories correspond to rational, trigonometric and elliptic Bethe ansatz equations, respectively. The initial partition function and effective superpotential of 3d $\mathcal{N}=2$ gauge theory on $D^{2}\times S^{1}$ has been given in \cite{YS20}. The Bethe ansatz equation of open $\text{XXZ}$ spin chain with diagonal boundary condition has been calculated in \cite{KZ21}, but for the case of $\text{Sp}$ gauge theory, the correspondence is not so directly, while the 2d degeneration can reproduce the duality with diagonal boundary condition $\text{XXX}$ spin chain. To cure this problem, in \cite{DZ23} we change the representation of gauge groups and adding weighted factors of root of Lie algebras into the effective superpotential $W_{\text{eff}}^{3d}(\sigma,m)$. With this variation, we can use a unified framework to deal gauge theory duality with open and closed cases. We got the square root of the new vacuum equations and regarded them as two independent connecting branches in the moduli space of vacuum of the correspondence gauge theory. From Lie theory, $\text{ADE}$ series are self-dual, and $B_n$ and $C_n$ Langlands dual to each other. The Bethe equation for a given boundary parameter matches only one branch of the squared root of the vacuum equation. In this way, the two branches for $\text{ADE}$ are the same. But for $B_n$ and $C_n$ case, they belong to different branches. In a certain scene, we deal with a double covering theory. From spin chain theories side along, our choice seems redundant. But, without this, we cannot treat $B_n$ and $C_n$ gauge theory equally. On the opposite view, we can view our results as to glue two certain spin chains to get a supersymmetric gauge theory. This is a new way to get a supersymmetric gauge theory. Then we gave a new Bethe/Gauge correspondence between 3d $A$, or $\text{BCD}$-type gauge theories and closed, open $\text{XXZ}$ spin chains with diagonal boundary conditions, respectively. As well as between 2d $\text{A}$, or $\text{BCD}$-type gauge theories and closed, open $\text{XXX}$ spin chains with diagonal boundary conditions, respectively.

For a 4d $\mathcal{N}=1$ supersymmetric gauge theory, the Seiberg duality \cite{S95} is a version of electric-magnetic duality in supersymmetric gauge theory.  For supersymmetric QCD, it identifies in the infrared the quarks and gluons in a theory with $N_f$ quark flavors and $\text{SU}(N_c=N)$ gauge group for $N_f-N\ >1$ with solitons in a theory of $N_f$ quark flavors and $\text{SU}(N_f-N)$ gauge group. The electric-magnetic duality, or the GNO duality  exchange the role of the simple roots and  fundamental weights of a Lie algebra \cite{GNO77}. In a special case, $N_f= 2N$, with all the quiver nodes having a gauge group with the same rank $N$. One advantage of this choice is that the rank of the gauge group, and hence the number of the components of the spin of the integrable model at a lattice site, is preserved by the Seiberg duality. The 4d $\mathcal{N}=1$ Seiberg dualities will be inherited as the duality between two supersymmetric quiver gauge theories, as it be realized as the duality between different solutions of the Yang-Baxter equation \cite{YY15}, and the quiver theory can be obtained by glue three $R$-matrices through a product of three $R$-matrices. 

Quivers, gauge theories and singular geometries are of great interest in both mathematics and physics. There are many questions which have arisen in various recent works at the intersection between gauge theories, representation theory, and algebraic geometry. There has been a lot of new progress in quiver gauge theory in recent years \cite{NPS18,KZ19,NW21}. The partition function of 4d $\mathcal{N}=2$, 5d $\mathcal{N}=1$ quiver gauge theory has been worked out. In general, the correspondence between gauge theory and spin chain is promoted to the higher rank cases \cite{CDHL11,CHZ12,LH11,NP12,NPS18}. 3d non-simply laced quiver gauge theory is  constructed in \cite{CK18}, as well as the Seiberg-like dualities in (2+1)d quiver gauge theories are considered in \cite{BGM21}. The Bethe/Gauge correspondence between 2d $A_{N}$ quiver gauge theory and $sl_{N+1}$ closed $\text{XXZ}$ spin chain has been given in \cite{NS09a}. And the Bethe ansatz equation of $sl_{3}$ $\text{XXZ}$ spin chain has been calculated in \cite{SP18,KZ21}. For 2d, 3d $A_{2}$ quiver gauge theories, the correspondence has been proved partly in \cite{KZ21}. The correspondence worked perfectly for 2d quiver gauge theories, but not as well in the 3d case. Similar to the case of $A_{1}$ quiver gauge theory of $\text{Sp}$ gauge group, the barrier is that the vacuum equation of 3d $\text{C}$-type gauge theory does not directly correspond to the Bethe ansatz equation of open $\text{XXZ}$ spin chain for a factor $\text{sin}^2(2\sigma_{i}\pm \beta_{2}\tilde{c})$ appearing in the vacuum equation.

In this article, following the correspondence between 2d, 3d gauge theory and $\text{XXZ}$ spin chain, we consider $A_{2}$ quiver gauge theory and $sl_{3}$ open $\text{XXZ}$ spin chain. For each gauge node in quiver gauge theories, we consider the adjoint chiral multiplet, fundamental multiplet and anti-fundamental multiplet. For each edge between two gauge nodes, we consider the bifundamental matter multiplet. As an example, we calculate the effective superpotential and the vacuum equations of $A_{2}$ quiver gauge theory with different product gauge groups. In particular, we only consider classical Lie group in this paper. More important, we find that the 2d correspondence between $A_{2}$ quiver gauge theory and $sl_{3}$ open $\text{XXZ}$ spin chain in \cite{KZ21} is a special circumstance of our results. We also calculate the vacuum equations of general $A_{r}$ quiver gauge theory with different type product gauge groups.

This article is organized as follows. In section \ref{a}, we give a brief introduction to the $sl_{3}$ $\text{XXZ}$ spin chain. In section \ref{b}, we define a new effective potential of the 3d $A_{2}$ quiver gauge theory. By using the effective potential, we first reproduce the Bethe/Gauge correspondence between 3d $A_{2}$ quiver gauge theory with product gauge group $\text{SU}(N_{1})\times \text{SU}(N_{2})$ and $sl_{3}$ $\text{XXX}$ spin chain with periodic boundary condition. We extend this duality to the 3d (or 2d) $A_{2}$ quiver gauge theory with $\text{BCD}$-type product gauge group and $sl_{3}$ open $\text{XXZ}$ (or $\text{XXX}$) spin chain with diagonal boundary condition in section \ref{c}. We further carry out the vacuum equation of general $A_{r}$ quiver gauge theory with classical Lie groups in section \ref{d}. Then we conclude this article and discuss the future work in section \ref{e}.

\section{$sl_3$ spin chain}\label{a}

The integrability of a spin chain is characterized by an $R$-matrix, $R(u):V\otimes V \rightarrow V\otimes V$. The $R$-matrix associated to 
the fundamental representation of a quantum group $U_{q}(\hat{sl_3})$ is known to take the form \cite{SP18}
\begin{equation}\label{1}
R(u)=\begin{pmatrix}
[u+\eta]&0&0&0&0&0&0&0&0\\
0&[u]&0&e^{i\pi u}[\eta]&0&0&0&0&0\\
0&0&[u]&0&0&0&e^{i\pi u}[\eta]&0&0\\
0&e^{-i\pi u}[\eta]&0&[u]&0&0&0&0&0\\
0&0&0&0&[u+\eta]&0&0&0&0\\
0&0&0&0&0&[u]&0&e^{i\pi u}[\eta]&0\\
0&0&e^{-i\pi u}[\eta]&0&0&0&[u]&0&0\\
0&0&0&0&0&e^{-i\pi u}[\eta]&0&[u]&0\\
0&0&0&0&0&0&0&0&[u+\eta]
\end{pmatrix}
\end{equation}
in the convention of this article, where 
\begin{equation}
    [x]=\dfrac{\text{sin}(\pi x)}{\text{sin}(\pi \eta)}
\end{equation}
The $R$-matrix processes the following properties,
\begin{itemize}
\item Initial condition:\quad $R_{12}(0)=P_{12}$.
\item Unitarity relation:\quad $R_{12}(u)R_{21}(-u)=-\text{sin}(u-\eta)\text{sin}(u+\eta)\times \mathbf{\text{id}}$.
\item Crossing Unitarity relation:\quad $R_{12}^{t_{1}}(u)\mathcal{M}_{1}R_{21}^{t_{1}}(-u-3\eta)\mathcal{M}_{1}^{-1}=-\text{sin}(u)\text{sin}(u+3\eta)\times \mathbf{\text{id}}$
\end{itemize}
where $\mathcal{M}$ are crossing matrix
\begin{equation}\label{4}
 \mathcal{M}=\begin{pmatrix}
 e^{4\eta}&0&0\\
 0&e^{2 \eta}&0\\
 0&0&1   
 \end{pmatrix}
\end{equation}
The $R$-matrix satisfies the quantum Yang-Baxter equation
\begin{equation}
    R_{12}(u_{1}-u_{2})R_{13}(u_{1}-u_{3})R_{23}(u_{2}-u_{3})=R_{23}(u_{2}-u_{3})R_{13}(u_{1}-u_{3})R_{12}(u_{1}-u_{2})
\end{equation}
The Bethe ansatz equation of a general periodic spin chain associated to the 
$R$-matrix of the $\text{Lie}$ algebra $\mathfrak{g}$ is well known in the literature \cite{DDMST07}. In the case of $\mathfrak{g}=sl_{3}$, there are two sets of Bethe roots, $\{u_{i}^{(1)}\}$ and $\{u_{i}^{(2)}\}$. The number of spin sites $L$ and the excitation level $N$ of our previous models generalize to the vectors: $L=(L_{1},L_{2})$ and $N=(N_{1},N_{2})$. The Bethe ansatz equations are \cite{BVV82,BVV83}
\begin{equation}\label{16}
    \prod_{a=1}^{L_{1}}\dfrac{[u_{i}^{(1)}-\theta_{a}^{(1)}+\frac{\eta}{2}]}{[u_{i}^{(1)}-\theta_{a}^{(1)}-\frac{\eta}{2}]}\prod_{j=1}^{N_{2}}\dfrac{[u_{i}^{(1)}-u_{i}^{(2)}+\frac{\eta}{2}]}{[u_{i}^{(1)}-u_{i}^{(2)}-\frac{\eta}{2}]}=\prod_{j=1}^{N_{1}}\dfrac{[u_{i}^{(1)}-u_{i}^{(1)}+\eta]}{[u_{i}^{(1)}-u_{i}^{(1)}-\eta]}
\end{equation}
and
\begin{equation}\label{17}
    \prod_{a=1}^{L_{2}}\dfrac{[u_{i}^{(2)}-\theta_{a}^{(2)}+\frac{\eta}{2}]}{[u_{i}^{(2)}-\theta_{a}^{(2)}-\frac{\eta}{2}]}\prod_{j=1}^{N_{1}}\dfrac{[u_{i}^{(2)}-u_{i}^{(1)}+\frac{\eta}{2}]}{[u_{i}^{(2)}-u_{i}^{(1)}-\frac{\eta}{2}]}=\prod_{j=1}^{N_{2}}\dfrac{[u_{i}^{(2)}-u_{i}^{(2)}+\eta]}{[u_{i}^{(2)}-u_{i}^{(2)}-\eta]}
\end{equation}
respectively. We focus on the open spin chain with diagonal boundary condition. Let us introduce the reflection matrix $K_{-}(u)$ and the dual one $K_{+}(u)$. 
\begin{equation}\label{2}
K_{-}(u)=\begin{pmatrix}
-e^{i\pi u}[u-\xi_{-}]&0&0\\
0&-e^{i\pi u}[u-\xi_{-}]&0\\
0&0&e^{-i \pi u}[u+\xi_{-}]
\end{pmatrix}
\end{equation}
and its dual 
\begin{equation}\label{3}
\begin{aligned}
K_{+}(u)&=\mathcal{M}K_{-}(-u-\frac{3}{2}\eta)\\
&=\begin{pmatrix}
e^{-i\pi u+\frac{5}{2}i\pi \eta}[u+\xi_{+}+\frac{3\eta}{2}]&0&0\\
0&e^{-i\pi u+\frac{1}{2}i\pi \eta}[u+\xi_{+}+\frac{3\eta}{2}]&0\\
0&0&-e^{i \pi u+\frac{3}{2}i\pi \eta}[u-\xi_{+}+\frac{3\eta}{2}]
\end{pmatrix}
\end{aligned}
\end{equation}
The reflection matrix $K_{-}(u)$ satisfies the reflection equation
\begin{equation}
\begin{aligned}
    &R_{12}(u_{1}-u_{2})K_{-}^{1}(u_{1})R_{12}(u_{1}+u_{2})K_{-}^{2}(u_{2})\\
    &=K_{-}^{2}(u_{2})R_{12}(u_{1}+u_{2})K_{-}^{1}(u_{1})R_{12}(u_{1}-u_{2})
\end{aligned}
\end{equation}
and the dual reflection matrix $K_{+}(u)$ satisfies the dual reflection equation
\begin{equation}
\begin{aligned}
    &R_{12}(u_{1}-u_{2})K_{+}^{1}(u_{1})\mathcal{M}_{1}^{-1}R_{12}(u_{1}+u_{2})\mathcal{M}_{1}K_{+}^{2}(u_{2})\\
    &=K_{+}^{2}\mathcal{M}_{2}^{-1}(u_{2})R_{12}(u_{1}+u_{2})\mathcal{M}_{2}K_{+}^{1}(u_{1})R_{12}(u_{1}-u_{2})
\end{aligned}
\end{equation}
In order to show the integrability of the system, we introduce the row-to-row monodromy matrices $T_{0}(u)$ and $\hat{T}_{0}(u)$
\begin{equation}
    T_{0}(u)=R_{0N}(u-\theta_{N})\cdots R_{01}(u-\theta_{1})
\end{equation}
\begin{equation}
     \hat{T}_{0}(u)=R_{10}(u+\theta_{1})\cdots R_{N0}(u+\theta_{N})
\end{equation}
where $\{\theta_{1},\cdots,\theta_{N}\}$ are the inhomogeneous parameters and $N$ is the number of sites. For an open spin chain, we need to define the double-row monodromy matrix $\mathbf{T}_{0}(u)$
\begin{equation}
    \mathbf{T}_{0}(u)=T_{0}K_{-}^{0}\hat{T}_{0}(u):=\begin{pmatrix}
A(u)&B_{1}(u)&B_{2}(u)\\
C_{1}(u)&D_{12}(u)&D_{12}(u)\\
C_{2}(u)&D_{21}(u)&D_{22}(u)
\end{pmatrix}
\end{equation}
Then the transfer matrix $t(u)$ can be construct as
\begin{equation}
    t(u)=tr_{0}\left(K_{+}^{0}\mathbf{T}_{0}(u)\right)
\end{equation}
The Bethe ansatz equations are \cite{SP18,KZ21}
\begin{equation}\label{5}
\begin{aligned}
    &\dfrac{[2u_{i}^{(1)}-\eta][u_{i}^{(1)}+\xi_{-}+\frac{\eta}{2}][u_{i}^{(1)}-\xi_{+}]}{[2u_{i}^{(1)}+\eta][u_{i}^{(1)}-\xi_{-}-\frac{\eta}{2}][u_{i}^{(1)}+\xi_{+}]}\prod_{j=1}^{N_{1}}\dfrac{[u_{i}^{(1)}-u_{j}^{(1)}-\eta][u_{i}^{(1)}+u_{j}^{(1)}-\eta]}{[u_{i}^{(1)}-u_{j}^{(1)}+\eta][u_{i}^{(1)}+u_{j}^{(1)}+\eta]}\\
    &\times \prod_{k=1}^{N_{2}}\dfrac{[u_{i}^{(1)}-u_{k}^{(2)}-\frac{\eta}{2}][u_{i}^{(1)}+u_{k}^{(2)}-\frac{\eta}{2}]}{[u_{i}^{(1)}-u_{k}^{(2)}+\frac{\eta}{2}][u_{i}^{(1)}+u_{k}^{(2)}+\frac{\eta}{2}]}\prod_{a=1}^{L_{1}}\dfrac{[u_{i}^{(1)}+\theta_{a}-\frac{\eta}{2}][u_{i}^{(1)}-\theta_{a}-\frac{\eta}{2}]}{[u_{i}^{(1)}+\theta_{a}+\frac{\eta}{2}][u_{i}^{(1)}-\theta_{a}+\frac{\eta}{2}]}=1
\end{aligned}
\end{equation}
and 
\begin{equation}\label{6}
\begin{aligned}
    &\dfrac{[2u_{i}^{(2)}+\eta][u_{i}^{(2)}+\xi_{-}][u_{i}^{(2)}-\xi_{+}-\frac{\eta}{2}]}{[2u_{i}^{(2)}-\eta][u_{i}^{(2)}-\xi_{-}][u_{i}^{(2)}+\xi_{+}+\frac{\eta}{2}]}\prod_{j=1}^{N_{2}}\dfrac{[u_{i}^{(2)}-u_{j}^{(2)}+\eta][u_{i}^{(2)}+u_{j}^{(2)}+\eta]}{[u_{i}^{(2)}-u_{j}^{(2)}-\eta][u_{i}^{(2)}+u_{j}^{(2)}-\eta]}\\
    &\times\prod_{k=1}^{N_{1}}\dfrac{[u_{i}^{(2)}+u_{k}^{(1)}+\frac{\eta}{2}][u_{i}^{(2)}-u_{k}^{(1)}+\frac{\eta}{2}]}{[u_{i}^{(2)}-u_{k}^{(1)}-\frac{\eta}{2}][u_{i}^{(2)}+u_{k}^{(1)}-\frac{\eta}{2}]}=1
\end{aligned}
\end{equation}
respectively.

\section{$A_2$ quiver gauge theory}\label{b}

The duality between 2d $A_{r}$ quiver gauge theory with gauge groups $G=U(N_{1})\times \cdots \times U(N_{r})$ and $sl_{r+1}$ $\text{XXX}$ spin chain with twisted periodic boundary condition is given in \cite{NS09a}. The spin operators $\vec{\mathbf{S}}_a$ are realized as the generators of some simple Lie
algebra $\mathbf{k}=\text{Lie}K$. Let $r=\text{rank}(\mathbf{k})$. The number of spin sites L and the excitation level $N$ of our previous models generalize to the vectors: $\vec{L}=(L_{1}, L_{2}, \cdots, L_{r})$ , $\Vec{N}=(N_{1},N_{2}, \cdots,N_{r})$. The twist parameter becomes the $r$-tuple of angles: $(\vartheta_{1},\cdots, \vartheta_{r})$, which
define an element of the maximal torus of $K$. The Bethe equations read as follows
\begin{equation}\label{9}
    \prod_{a=1}^{L_{\mathbf{i}}}\dfrac{\lambda_{i}^{(\mathbf{i})}-\theta_{a}^{(\mathbf{i})}+is_{a}^{(\mathbf{i})}}{\lambda_{i}^{(\mathbf{i})}-\theta_{a}^{(\mathbf{i})}-is_{a}^{(\mathbf{i})}}=e^{i\vartheta_{i}}\prod_{\mathbf{j}=1}^{r}\prod_{j:(i,\mathbf{i})\neq (j,\mathbf{j})}\dfrac{\lambda_{i}^{(\mathbf{i})}-\lambda_{j}^{(\mathbf{j})}+\frac{i}{2}\mathcal{C}_{\mathbf{ij}}}{\lambda_{i}^{(\mathbf{i})}-\lambda_{j}^{(\mathbf{j})}-\frac{i}{2}\mathcal{C}_{\mathbf{ij}}}
\end{equation}
where the unknown Bethe roots are $\lambda_{i}^{(i)}$, $\mathbf{i}=1,\cdots,r$, $i=1,\cdots,N_{\mathbf{i}} $. The above Bethe equations describe the spectrum of the transfer matrix acting in the space
\begin{equation}
    \mathcal{H}_{\Vec{L}}=\bigotimes_{\mathbf{i}=1}^{r}\otimes _{a=1}^{L_{\mathbf{i}}}\mathcal{W}_{s_{a}^{\mathbf{(i)}}}^{\mathbf{(i)}}\left(\theta_{a}^{\mathbf{(i)}}\right)
\end{equation}
where $\mathcal{W}_{s}^{\mathbf{i}}(\theta)$, $2s\in \mathbf{Z}_{\ge 0}$, $\theta \in \mathbf{C}$ are the so-called Kirillov-Reshetikhin modules \cite{KR90}, the special evaluation representations of the Yangian $\mathcal{Y}(\mathbf{k})$ of $\mathbf{k}$. The matrix $\mathcal{C}_{\mathbf{ij}}$ in (\ref{9}) is the Cartan matrix of $\mathbf{k}$. For simple laced Lie algebras, the Cartan matrices $\mathcal{C}_{\mathbf{ij}}$ are symmetric, and we can use them to get Bethe equations directly. But for non-simple laced Lie algebras, their Cartan matrices are not symmetric, and we need to replace them with the symmetrized ones to get the right Bethe equations \cite{RW87}.

Now, we consider a 3d gauge theory with $A_{2}$ quiver gauge structure. For each gauge node with $\text{BCD}$-type gauge group
, the representation of the gauge group is 
\begin{equation}\label{101}
    \mathcal{R}=V\otimes V^{*} \oplus V\otimes \mathcal{F}\oplus V^{*}\otimes \tilde{\mathcal{F}}
\end{equation}
where $V$ is the standard representation of $\text{SU}(N)$, $V^{*}$ is the dual of $V$, $\mathcal{F}$ and $\Tilde{\mathcal{F}}$ are the fundamental representation and anti-fundamental representation respectively. The adjoint representation of $\text{SO}(M)$ is isomorphic to the wedge product $\bigwedge^{2} V$. In the case $M=2N$, since the weights of $V$ are $\pm e_{i}$, it follows that the roots of $\text{SO}(2N)$ are just the pairwise distinct sum $\pm e_{i}\pm e_j$.  In the odd case $M=2N+1$, the weights of the standard representation $V$ are $\{\pm e_{i}\} \cup \{0\}$ and the weights of adjoint representation are $\{\pm e_{i}\pm e_{j}\} \cup \{\pm e_{i}\}$. To make a comparison with the Lie algebra $\text{Sp}(2N)$, we can say that the root diagram of $\text{Sp}(2N)$ looks like that of $\text{SO}(2N)$ with the roots $\pm 2e_i$ removed, whereas the root diagram of $\text{Sp}(2N)$ looks like that of $\text{SO}(2N+1)$ with the roots $\pm 2e_i$ replaced by $\pm e_{i}$. See the book \cite{FH04} for details. The bifundamental representation is $\oplus_{2}(V_{1}\otimes V_{2})$, which means $(V_{1}\otimes V_{2}^{*})\oplus (V_{1}^{*}\otimes V_{2})$. Here $V_{1}$ is the fundamental representation of gauge group on the first node and $V_{2}$ is the fundamental representation of gauge group on the second node. The effective superpotential of each node comes from \cite{DZ23}
\begin{equation}\label{}
    \begin{aligned}
    W^{3d}_{\text{eff}}(\sigma,m)=&\dfrac{1}{\beta_{2}}\sum_{w \in \mathcal{R}}\sum_{a=1}^{N_{f}}\text{Li}_{2}(e^{-iw\cdot \sigma-im_{a}-i\beta_{2}\tilde{c}})-\dfrac{1}{4\beta_{2}}\sum_{w\in \mathcal{R}}\sum_{a=1}^{N_{f}}(w\cdot \sigma+m_{a}+\beta_{2}\tilde{c})^{2}\\
    &+\dfrac{1}{\beta_{2}}\sum_{w \in \mathcal{R}}\sum_{a=1}^{N'_{f}}\text{Li}_{2}(e^{iw\cdot \sigma-im'_{a}-i\beta_{2}\tilde{c}})-\dfrac{1}{4\beta_{2}}\sum_{w\in \mathcal{R}}\sum_{a=1}^{N'_{f}}(w\cdot \sigma-m'_{a}+\beta_{2}\tilde{c})^{2}\\
    &-\dfrac{1}{\beta^{2}}\sum_{\alpha \in \Delta}\frac{4}{\alpha_{i}^{2}}\text{Li}_{2}(e^{i\alpha\cdot \sigma})+\dfrac{1}{4\beta_{2}}\sum_{\alpha\in \Delta}\frac{4}{\alpha_{i}^{2}}(\alpha\cdot \sigma)^{2}
   \end{aligned}
\end{equation}
where the weight's set $w$ of the corresponding representation denoted by $\mathcal{R}$, and $\sigma_{i}$ are the eigenvalues of the complex scalar in the vector multiplet, $\beta_{2}$ is the $U(1)$ charge fugacity, $\tilde{c}$ is the rescaling $\text{R}$-charge of the scalar in the chiral multiplet, ${m_a}$ is the twisted mass parameter associated with a given representation. The set of the roots of $G$ is denoted by $\Delta$, and $\alpha_{i}$ are roots of the Lie algebra 
$\mathfrak{g}$ of the Lie groups $G$, respectively. The $\text{Li}_{2}(z)$ is called the dilogarithm function
\begin{equation}
\text{Li}_{2}(z)=\sum_{k=1}^{\infty}\dfrac{z^{k}}{k^{2}}
\end{equation}
According to our assumptions, we have $N_{f}=N'_{f}$. But it does not mean that $N_{f}^{(1)}=N_{f}^{(2)}$. Only for every gauge node with $\text{SU}(N)$-type gauge group, we can set the representation (\ref{101}) with twice fundamental multiplets and twice anti-fundamental multiplets
\begin{equation}\label{102}
    \mathcal{R}=V\otimes V^{*}\oplus V\otimes \mathcal{F}\oplus V\otimes \mathcal{F}\oplus V^{*}\otimes \tilde{\mathcal{F}}\oplus V^{*}\otimes \tilde{\mathcal{F}}
\end{equation}
to consist with the following $A_2$ quiver gauge theory with $\text{BCD}$-type gauge group. The contribution of the four bifundamental chiral multiplets to the effective superpotential is 
\begin{equation}
    \begin{aligned}
    W^{3d, \text{bfd}}_{\text{eff}}=&\dfrac{2}{\beta_{2}}\sum_{j=1}^{N_{1}}\sum_{k=1}^{N_{2}}\text{Li}_{2}(e^{-i(\sigma_{j}^{(1)}- \sigma_{k}^{(2)})-im_{\text{bfd}}})-\dfrac{1}{2\beta_{2}}\sum_{j=1}^{N_{1}}\sum_{k=1}^{N_{2}}( \sigma_{j}^{(1)}- \sigma_{k}^{(2)}+m_{\text{bfd}})^{2}\\
    &+\dfrac{2}{\beta_{2}}\sum_{j=1}^{N_{2}}\sum_{k=1}^{N_{1}}\text{Li}_{2}(e^{-i(\sigma_{j}^{(2)}- \sigma_{k}^{(1)})-im_{\text{bfd}}})-\dfrac{1}{2\beta_{2}}\sum_{j=1}^{N_{2}}\sum_{k=1}^{N_{1}}( \sigma_{j}^{(2)}-\sigma_{k}^{(1)}+m_{\text{bfd}})^{2}
    \end{aligned}
\end{equation}
The vacuum equations of the $\text{SU}(N_{1})\times \text{SU}(N_{2})$ quiver gauge theory are
\begin{equation}\label{18}
\prod_{k\neq j}^{N_{1}}\dfrac{\text{sin}^{2}(\sigma_{j}^{(1)}-\sigma_{k}^{(1)}-m_{\text{adj}}^{(1)})}{\text{sin}^{2}(\sigma_{j}^{(1)}-\sigma_{k}^{(1)}+m_{\text{adj}}^{(1)})}
\dfrac{\prod_{a=1}^{N'_{f}}\text{sin}^{2}(\sigma_{j}^{(1)}-{m'}_{a}^{(1)})}{\prod_{a=1}^{N_{f}}\text{sin}^{2}(\sigma_{j}^{(1)}+m_{a}^{(1)})}
\prod_{l=1}^{N_{2}}\dfrac{\text{sin}^{2}(\sigma_{l}^{(2)}-\sigma_{j}^{(1)}+m_{\text{bfd}})}{\text{sin}^{2}(\sigma_{l}^{(1)}-\sigma_{j}^{(2)}+m_{\text{bfd}})}=1
\end{equation}
and
\begin{equation}\label{19}
\prod_{k\neq j}^{N_{2}}\dfrac{\text{sin}^{2}(\sigma_{j}^{(2)}-\sigma_{k}^{(2)}-m_{\text{adj}}^{(2)})}{\text{sin}^{2}(\sigma_{j}^{(2)}-\sigma_{k}^{(2)}+m_{\text{adj}}^{(2)})}
\dfrac{\prod_{a=1}^{N'_{f}}\text{sin}^{2}(\sigma_{j}^{(2)}-{m'}_{a}^{(2)})}{\prod_{a=1}^{N_{f}}\text{sin}^{2}(\sigma_{j}^{(2)}+m_{a}^{(2)})}
\prod_{l=1}^{N_{1}}\dfrac{\text{sin}^{2}(\sigma_{l}^{(1)}-\sigma_{j}^{(2)}+m_{\text{bfd}})}{\text{sin}^{2}(\sigma_{l}^{(2)}-\sigma_{j}^{(1)}+m_{\text{bfd}})}=1
\end{equation}
where we have absorbed $\beta_{2}\tilde{c}$ into the twisted mass parameter $m_{a}$ and the adjoint mass parameter $m_{\text{adj}}$. We can get two genres of vacuum equations after taking the square root of (\ref{18})
\begin{equation}\label{20}
\prod_{k\neq j}^{N_{1}}\dfrac{\text{sin}(\sigma_{j}^{(1)}-\sigma_{k}^{(1)}-m_{\text{adj}}^{(1)})}{\text{sin}(\sigma_{j}^{(1)}-\sigma_{k}^{(1)}+m_{\text{adj}}^{(1)})}
\dfrac{\prod_{a=1}^{N'_{f}}\text{sin}(\sigma_{j}^{(1)}-{m'}_{a}^{(1)})}{\prod_{a=1}^{N_{f}}\text{sin}(\sigma_{j}^{(1)}+m_{a}^{(1)})}
\prod_{l=1}^{N_{2}}\dfrac{\text{sin}(\sigma_{l}^{(2)}-\sigma_{j}^{(1)}+m_{\text{bfd}})}{\text{sin}(\sigma_{l}^{(1)}-\sigma_{j}^{(2)}+m_{\text{bfd}})}=1
\end{equation}
and
\begin{equation}\label{}
\prod_{k\neq j}^{N_{1}}\dfrac{\text{sin}(\sigma_{j}^{(1)}-\sigma_{k}^{(1)}-m_{\text{adj}}^{(1)})}{\text{sin}(\sigma_{j}^{(1)}-\sigma_{k}^{(1)}+m_{\text{adj}}^{(1)})}
\dfrac{\prod_{a=1}^{N'_{f}}\text{sin}(\sigma_{j}^{(1)}-{m'}_{a}^{(1)})}{\prod_{a=1}^{N_{f}}\text{sin}(\sigma_{j}^{(1)}+m_{a}^{(1)})}
\prod_{l=1}^{N_{2}}\dfrac{\text{sin}(\sigma_{l}^{(2)}-\sigma_{j}^{(1)}+m_{\text{bfd}})}{\text{sin}(\sigma_{l}^{(1)}-\sigma_{j}^{(2)}+m_{\text{bfd}})}=-1
\end{equation}
Similarly, we get the vacuum equations after taking the square root of (\ref{19})
\begin{equation}\label{21}
\prod_{k\neq j}^{N_{2}}\dfrac{\text{sin}(\sigma_{j}^{(2)}-\sigma_{k}^{(2)}-m_{\text{adj}}^{(2)})}{\text{sin}(\sigma_{j}^{(2)}-\sigma_{k}^{(2)}+m_{\text{adj}}^{(2)})}
\dfrac{\prod_{a=1}^{N'_{f}}\text{sin}(\sigma_{j}^{(2)}-{m'}_{a}^{(2)})}{\prod_{a=1}^{N_{f}}\text{sin}(\sigma_{j}^{(2)}+m_{a}^{(2)})}
\prod_{l=1}^{N_{1}}\dfrac{\text{sin}(\sigma_{l}^{(1)}-\sigma_{j}^{(2)}+m_{\text{bfd}})}{\text{sin}(\sigma_{l}^{(2)}-\sigma_{j}^{(1)}+m_{\text{bfd}})}=1
\end{equation}
and
\begin{equation}\label{}
\prod_{k\neq j}^{N_{2}}\dfrac{\text{sin}(\sigma_{j}^{(2)}-\sigma_{k}^{(2)}-m_{\text{adj}}^{(2)})}{\text{sin}(\sigma_{j}^{(2)}-\sigma_{k}^{(2)}+m_{\text{adj}}^{(2)})}
\dfrac{\prod_{a=1}^{N'_{f}}\text{sin}(\sigma_{j}^{(2)}-{m'}_{a}^{(2)})}{\prod_{a=1}^{N_{f}}\text{sin}(\sigma_{j}^{(2)}+m_{a}^{(2)})}
\prod_{l=1}^{N_{1}}\dfrac{\text{sin}(\sigma_{l}^{(1)}-\sigma_{j}^{(2)}+m_{\text{bfd}})}{\text{sin}(\sigma_{l}^{(2)}-\sigma_{j}^{(1)}+m_{\text{bfd}})}=-1
\end{equation}
We can equal the vacuum equation (\ref{20}) with the Bethe ansatz equation (\ref{16}), and match the vacuum equation (\ref{21}) with the Bethe ansatz equation (\ref{17}). The dictionary is given by
\begin{equation}
    \begin{aligned}
    &\pi \eta \longleftrightarrow m_{\text{adj}}^{(1)}=m_{\text{adj}}^{(2)},\quad -\dfrac{\pi \eta}{2}\longleftrightarrow m_\text{bfd}\\
    &-\pi\theta_{a}^{(i)}-\pi\dfrac{\eta}{2}\longleftrightarrow m_{a}^{(i)},\quad \pi\theta_{a}^{(i)}-\pi\frac{\eta}{2}\longleftrightarrow m_{a}^{'(i)}
\end{aligned}
\end{equation}
The correspondence here certainly works in parallel after taking 2d limit, $\text{sin}\sigma\rightarrow \sigma$, in the quiver gauge theory and the $\text{XXX}$ limit, $[u]\rightarrow u$, of the spin chain.

Actually, for quiver theory at every node with $\text{SU}(N)$ gauge group, the representation (\ref{101}) is enough to get the vacuum equation given by \cite{NS09a}. But if we consider quiver theory with gauge group $\text{SU}(N)\times \text{SO}(M)$, or $\text{SU}(N)\times \text{Sp}(2M)$, the bifundamental multiplets will be tricky to deal with. Using the representation (\ref{102}), we can take square root for all kinds of quiver gauge groups. And this problem is left for future publication.

\section{Bethe/Gauge correspondence}\label{c}
In this section, we compute the effective superpotential and the vacuum equations of $A_2$ quiver gauge theories with $\text{BCD}$-type product gauge groups. We explore the correspondence between $A_2$ quiver gauge theory and $sl_3$ open spin chain model. 

\subsection{$\text{SO}(2N_1+1)\times \text{SO}(2N_2+1)$}

For $\text{SO}(2N_{1}+1)\times \text{SO}(2N_{2}+1)$ quiver gauge theory, i.e. one gauge node (say the first node) is $\text{SO}(2N_{1}+1)$ gauge group and the other (the second node) is $\text{SO}(2N_{2}+1)$ gauge group, we glue two gauge nodes with two bifundamental chiral multiplets. The effective potential of the bifundamental multiplets is given by
\begin{equation}
    \begin{aligned}
    W^{3d, \text{bfd}}_{\text{eff}}=&\dfrac{1}{\beta_{2}}\sum_{j=1}^{N_{1}}\sum_{k=1}^{N_{2}}\text{Li}_{2}(e^{-i(\pm\sigma_{j}^{(1)}\pm \sigma_{k}^{(2)})-im_{\text{bfd}}})-\dfrac{1}{4\beta_{2}}\sum_{j=1}^{N_{1}}\sum_{k=1}^{N_{2}}(\pm \sigma_{j}^{(1)}\pm \sigma_{k}^{(2)}+m_{\text{bfd}})^{2}\\
    &+\dfrac{1}{\beta_{2}}\sum_{j=1}^{N_{2}}\sum_{k=1}^{N_{1}}\text{Li}_{2}(e^{-i(\pm\sigma_{j}^{(2)}\pm \sigma_{k}^{(1)})-im_{\text{bfd}}})-\dfrac{1}{4\beta_{2}}\sum_{j=1}^{N_{2}}\sum_{k=1}^{N_{1}}(\pm \sigma_{j}^{(2)}\pm \sigma_{k}^{(1)}+m_{\text{bfd}})^{2}\\
    &+\dfrac{2}{\beta_{2}}\sum_{j=1}^{N_{1}}\text{Li}_{2}(e^{-i(\pm\sigma_{j}^{(1)})-im_{\text{bfd}}})-\dfrac{1}{2\beta_{2}}\sum_{j=1}^{N_{1}}(\pm\sigma_{j}^{(1)}+m_{\text{bfd}})^{2}\\
    &+\dfrac{2}{\beta_{2}}\sum_{j=1}^{N_{2}}\text{Li}_{2}(e^{-i(\pm\sigma_{j}^{(2)})-im_{\text{bfd}}})-\dfrac{1}{2\beta_{2}}\sum_{j=1}^{N_{2}}(\pm\sigma_{j}^{(2)}+m_{\text{bfd}})^{2}
    \end{aligned}
\end{equation}
For the first node, the effective superpotential is
\begin{equation}\label{}
\begin{aligned}
W^{3d}_{\text{eff}}(\sigma,m)=&-\dfrac{2}{\beta_{2}}\sum_{j<k}^{N_{1}}\text{Li}_{2}(e^{i(\pm\sigma_{j}^{(1)}\pm \sigma_{k}^{(1)})})+\dfrac{1}{2\beta_{2}}\sum_{j<k}^{N_{1}}(\pm \sigma_{j}^{(1)}\pm \sigma_{k}^{(1)})^{2}\\
&+\dfrac{2}{\beta_{2}}\sum_{j<k}^{N_{1}}\text{Li}_{2}(e^{-i(\pm \sigma_{j}^{(1)}\pm \sigma_{k}^{(1)})-im_{adj}^{(1)}})-\dfrac{1}{2\beta_{2}}\sum_{j<k}^{N_{1}}(\pm \sigma_{j}^{(1)}\pm \sigma_{k}^{(1)}+m_{adj}^{(1)})^{2}\\
&+\dfrac{1}{\beta_{2}}\sum_{j=1}^{N_{1}}\sum_{a=1}^{N_{f}^{(1)}}\text{Li}_{2}(e^{-(\pm i \sigma_{j}^{(1)}+im_{a}^{(1)})})-\dfrac{1}{4\beta_{2}}\sum_{j=1}^{N_{1}}\sum_{a=1}^{N_{f}^{(1)}}(\pm \sigma_{j}^{(1)}+m_{a}^{(1)})^{2}\\
&-\dfrac{4}{\beta_{2}}\sum_{j=1}^{N_{1}}\text{Li}_{2}(e^{\pm i \sigma_{j}})+\dfrac{2}{\beta_{2}}\sigma_{j}^{2}\\
&+\dfrac{4}{\beta_{2}}\sum_{j=1}^{N_{1}}\text{Li}_{2}(e^{\pm i\sigma_{j}-im_{\text{adj}}^{(1)}})-\dfrac{1}{\beta_{2}}\sum_{j=1}^{N_{1}}(\sigma_{j}\pm m_{\text{adj}}^{(1)})^{2}\\
&+\dfrac{1}{\beta_{2}}\sum_{j=1}^{N_{1}}\sum_{a=1}^{{N'}^{(1)}_{f}}
\text{Li}_{2}(e^{\pm i \sigma_{j}^{(1)}-i{m'}_{a}^{(1)}})-\dfrac{1}{4\beta_{2}}\sum_{j=1}^{N_{1}}\sum_{a=1}^{{N'}_{f}^{(1)}}(\pm \sigma_{j}^{(1)}-{m'}_{a}^{(1)})^{2}\\
\qquad \qquad \qquad &+\dfrac{1}{\beta_{2}}\sum_{j=1}^{N_{1}}\sum_{k=1}^{N_{2}}\text{Li}_{2}(e^{-i(\pm\sigma_{j}^{(1)}\pm \sigma_{k}^{(2)})-i m_{\text{bfd}}})-\dfrac{1}{4\beta_{2}}\sum_{j=1}^{N_{1}}\sum_{k=1}^{N_{2}}(\pm \sigma_{j}^{(1)}\pm \sigma_{k}^{(2)}+m_{\text{bfd}})^{2}\\
&+\dfrac{1}{\beta_{2}}\sum_{j=1}^{N_{1}}\sum_{k=1}^{N_{2}}\text{Li}_{2}(e^{-i(\pm\sigma_{j}^{(1)}\pm \sigma_{k}^{(2)})-i m_{\text{bfd}}})-\dfrac{1}{4\beta_{2}}\sum_{j=1}^{N_{1}}\sum_{k=1}^{N_{2}}(\pm \sigma_{j}^{(1)}\pm \sigma_{k}^{(2)}+m_{\text{bfd}})^{2}\\
&+\dfrac{2}{\beta_{2}}\sum_{j=1}^{N_{1}}\text{Li}_{2}(e^{\pm i\sigma_{j}^{(1)}-im_{\text{bfd}}})-\dfrac{1}{2\beta_{2}}\sum_{j=1}^{N_{1}}(\sigma_{j}^{(1)}\pm m_{\text{bfd}})^{2}\\
&+\dfrac{2}{\beta_{2}}\sum_{j=1}^{N_{2}}\text{Li}_{2}(e^{\pm i\sigma_{j}^{(2)}-im_{\text{bfd}}})-\dfrac{1}{2\beta_{2}}\sum_{j=1}^{N_{2}}(\sigma_{j}^{(2)}\pm m_{\text{bfd}})^{2}
\end{aligned}
\end{equation}
The vacuum equation is given by
\begin{equation}\label{25}
\begin{aligned}
&\dfrac{\text{sin}^{4}(\sigma_{j}^{(1)}-m_{\text{adj}}^{(1)})}{\text{sin}^{4}(\sigma_{j}^{(1)}+m_{\text{adj}}^{(1)})}\prod_{j\neq k}^{N_{1}}\dfrac{\text{sin}^{2}(\sigma_{j}^{(1)}\pm \sigma_{k}^{(1)}-m_{adj}^{(1)})}{\text{sin}^{2}(-\sigma_{j}^{(1)}\pm \sigma_{k}^{(1)}-m_{adj}^{(1)})}\prod_{a=1}^{N_{f}^{(1)}}\dfrac{\text{sin}(\sigma_{j}^{(1)}-m_{a}^{(1)})}{\text{sin}(-\sigma_{j}^{(1)}-m_{a}^{(1)})}\\
&\prod_{a=1}^{{N'}_{f}^{(1)}}\dfrac{\text{sin}(\sigma_{j}^{(1)}-{m'}_{a}^{(1)})}{\text{sin}(-\sigma_{j}^{(1)}-{m'}_{a}^{(1)})}\prod_{k=1}^{N_{2}}\dfrac{\text{sin}^{2}(\sigma_{j}^{(1)}\pm \sigma_{k}^{(2)}-m_{\text{bfd}})}{\text{sin}^{2}(-\sigma_{j}^{(1)}\pm \sigma_{k}^{(2)}-m_{\text{bfd}})}\dfrac{\text{sin}^{2}(\sigma_{j}^{(1)}-m_{\text{bfd}})}{\text{sin}^{2}(\sigma_{j}^{(1)}+m_{\text{bfd}})}=1
\end{aligned}
\end{equation}
 For a gauge theory, $N_{f}^{(1)}$ and ${N'}_{f}^{(1)}$ , as well as 
$m_{a}^{(1)}$ and ${m'}_{a}^{(1)}$ are independent quantities. From the spin chain point of view, all quantities must be packing into a given spin chain. In this way, the independent quantities of quiver gauge theories are too much more for spin chain, and the Bethe/Gauge correspondence is out of focus. But, if we set $N_{f}^{(1)}={N'}_{f}^{(1)}$ and $m_{a}^{(1)}={m'}_{a}^{(1)}$, we can rewrite the vacuum equation with the square root of (\ref{25})
\begin{equation}\label{10}
    \begin{aligned}
        &\prod_{j\neq k}^{N_{1}}\dfrac{\text{sin}(\sigma_{j}^{(1)}\pm \sigma_{k}^{(1)}-m_{adj}^{(1)})}{\text{sin}(-\sigma_{j}^{(1)}\pm \sigma_{k}^{(1)}-m_{adj}^{(1)})}\prod_{a=1}^{N_{f}^{(1)}}\dfrac{\text{sin}(\sigma_{j}^{(1)}-m_{a}^{(1)})}{\text{sin}(-\sigma_{j}^{(1)}-m_{a}^{(1)})}\\
        &\times\prod_{k=1}^{N_{2}}\dfrac{\text{sin}(\sigma_{j}^{(1)}\pm \sigma_{k}^{(2)}-m_{\text{bfd}})}{\text{sin}(-\sigma_{j}^{(1)}\pm \sigma_{k}^{(2)}-m_{\text{bfd}})}\dfrac{\text{sin}^{2}(\sigma_{j}^{(1)}-m_{\text{adj}}^{(1)})}{\text{sin}^{2}(\sigma_{j}^{(1)}+m_{\text{adj}}^{(1)})}\dfrac{\text{sin}(\sigma_{j}^{(1)}-m_{\text{bfd}})}{\text{sin}^{2}(\sigma_{j}^{(1)}+m_{\text{bfd}})}=1
    \end{aligned}
\end{equation}
and
\begin{equation}\label{}
    \begin{aligned}
        &\prod_{j\neq k}^{N_{1}}\dfrac{\text{sin}(\sigma_{j}^{(1)}\pm \sigma_{k}^{(1)}-m_{adj}^{(1)})}{\text{sin}(-\sigma_{j}^{(1)}\pm \sigma_{k}^{(1)}-m_{adj}^{(1)})}\prod_{a=1}^{N_{f}^{(1)}}\dfrac{\text{sin}(\sigma_{j}^{(1)}-m_{a}^{(1)})}{\text{sin}(-\sigma_{j}^{(1)}-m_{a}^{(1)})}\\
        &\times\prod_{k=1}^{N_{2}}\dfrac{\text{sin}(\sigma_{j}^{(1)}\pm \sigma_{k}^{(2)}-m_{\text{bfd}})}{\text{sin}(-\sigma_{j}^{(1)}\pm \sigma_{k}^{(2)}-m_{\text{bfd}})}\dfrac{\text{sin}^{2}(\sigma_{j}^{(1)}-m_{\text{adj}}^{(1)})}{\text{sin}^{2}(\sigma_{j}^{(1)}+m_{\text{adj}}^{(1)})}\dfrac{\text{sin}(\sigma_{j}^{(1)}-m_{\text{bfd}})}{\text{sin}(\sigma_{j}^{(1)}+m_{\text{bfd}})}=-1
    \end{aligned}
\end{equation}
respectively. For the second node, the effective superpotential is 
\begin{equation*}\label{}
\begin{aligned}
W^{3d}_{\text{eff}}(\sigma,m)=&-\dfrac{2}{\beta_{2}}\sum_{j<k}^{N_{2}}\text{Li}_{2}(e^{i(\pm\sigma_{j}^{(2)}\pm \sigma_{k}^{(2)})})+\dfrac{1}{2\beta_{2}}\sum_{j<k}^{N_{2}}(\pm \sigma_{j}^{(2)}\pm \sigma_{k}^{(2)})^{2}\\
&+\dfrac{2}{\beta_{2}}\sum_{j<k}^{N_{2}}\text{Li}_{2}(e^{-i(\pm \sigma_{j}^{(2)}\pm \sigma_{k}^{(2)})-im_{adj}^{(2)}})-\dfrac{1}{2\beta_{2}}\sum_{j<k}^{N_{2}}(\pm \sigma_{j}^{(2)}\pm \sigma_{k}^{(2)}+m_{adj}^{(2)})^{2}\\
&+\dfrac{1}{\beta_{2}}\sum_{j=1}^{N_{2}}\sum_{a=1}^{N_{f}^{(2)}}\text{Li}_{2}(e^{-(\pm i \sigma_{j}^{(2)}+im_{a}^{(2)})})-\dfrac{1}{4\beta_{2}}\sum_{j=1}^{N_{2}}\sum_{a=1}^{N_{f}^{(2)}}(\pm \sigma_{j}^{(2)}+m_{a}^{(2)})^{2}\\
&-\dfrac{4}{\beta_{2}}\sum_{j=1}^{N_{2}}\text{Li}_{2}(e^{\pm i \sigma_{j}})+\dfrac{2}{\beta_{2}}\sigma_{2}^{2}\\
&+\dfrac{4}{\beta_{2}}\sum_{j=1}^{N_{2}}\text{Li}_{2}(e^{(\pm i\sigma_{j}-im_{\text{adj}}^{(2)})})-\dfrac{1}{\beta_{2}}\sum_{j=1}^{N_{1}}(\sigma_{j}\pm m_{\text{adj}}^{(2)})^{2}\\
&+\dfrac{1}{\beta_{2}}\sum_{j=1}^{N_{2}}\sum_{a=1}^{{N'}_{f}^{(2)}}\text{Li}_{2}(e^{\pm i \sigma_{j}^{(2)}-im_{a}^{(2)}})-\dfrac{1}{4\beta_{2}}\sum_{j=1}^{N_{2}}\sum_{a=1}^{{N'}_{f}^{(2)}}(\pm \sigma_{j}^{(2)}-m_{a}^{(2)})^{2}\\
&+\dfrac{1}{\beta_{2}}\sum_{j=1}^{N_{1}}\sum_{k=1}^{N_{2}}\text{Li}_{2}(e^{-i(\pm\sigma_{j}^{(1)}\pm \sigma_{k}^{(2)})-i m_{\text{bfd}}})-\dfrac{1}{2\beta_{2}}\sum_{j=1}^{N_{1}}\sum_{k=1}^{N_{2}}(\pm \sigma_{j}^{(1)}\pm \sigma_{k}^{(2)}+m_{\text{bfd}}))^{2}\\
&+\dfrac{1}{\beta_{2}}\sum_{j=1}^{N_{1}}\sum_{k=1}^{N_{2}}\text{Li}_{2}(e^{-i(\pm\sigma_{j}^{(1)}\pm \sigma_{k}^{(2)})-i m_{\text{bfd}}})-\dfrac{1}{4\beta_{2}}\sum_{j=1}^{N_{1}}\sum_{k=1}^{N_{2}}(\pm \sigma_{j}^{(1)}\pm \sigma_{k}^{(2)}+m_{\text{bfd}})^{2}\\
\end{aligned}
\end{equation*}
\begin{equation}
\begin{aligned}
&+\dfrac{2}{\beta_{2}}\sum_{j=1}^{N_{1}}\text{Li}_{2}(e^{\pm i\sigma_{j}^{(1)}-im_{\text{bfd}}})-\dfrac{1}{2\beta_{2}}\sum_{j=1}^{N_{1}}(\sigma_{j}^{(1)}\pm m_{\text{bfd}})^{2}\\
&+\dfrac{2}{\beta_{2}}\sum_{j=1}^{N_{2}}\text{Li}_{2}(e^{\pm i\sigma_{j}^{(2)}-im_{\text{bfd}}})-\dfrac{1}{2\beta_{2}}\sum_{j=1}^{N_{2}}(\sigma_{j}^{(2)}\pm m_{\text{bfd}})^{2}
\end{aligned}
\end{equation}
The vacuum equation is
\begin{equation}\label{26}
\begin{aligned}
&\dfrac{\text{sin}^{4}(\sigma_{j}^{(2)}-m_{\text{adj}}^{(2)})}{\text{sin}^{4}(\sigma_{j}^{(2)}+m_{\text{adj}}^{(2)})}\prod_{j\neq k}^{N_{2}}\dfrac{\text{sin}^{2}(\sigma_{j}^{(2)}\pm \sigma_{k}^{(2)}-m_{adj}^{(2)})}{\text{sin}^{2}(-\sigma_{j}^{(2)}\pm \sigma_{k}^{(2)}-m_{adj}^{(2)})}\prod_{a=1}^{N_{f}^{(2)}}\dfrac{\text{sin}(\sigma_{j}^{(2)}-m_{a}^{(2)})}{\text{sin}(-\sigma_{j}^{(2)}-m_{a}^{(2)})}\\
&\times\prod_{a=1}^{{N'}_{f}^{(2)}}\dfrac{\text{sin}(\sigma_{j}^{(2)}-{m'}_{a}^{(2)})}{\text{sin}(-\sigma_{j}^{(2)}-{m'}_{a}^{(2)})}\prod_{k=1}^{N_{1}}\dfrac{\text{sin}^{2}(\sigma_{j}^{(2)}\pm \sigma_{k}^{(1)}-m_{\text{bfd}})}{\text{sin}^{2}(-\sigma_{j}^{(2)}\pm \sigma_{k}^{(1)}-m_{\text{bfd}})}\dfrac{\text{sin}^{2}(\sigma_{j}^{(2)}-m_{\text{bfd}})}{\text{sin}^{2}(\sigma_{j}^{(2)}+m_{\text{bfd}})}=1
\end{aligned}
\end{equation}
We obtain two types of vacuum equation with the square root of (\ref{26}) with $N_{f}^{(2)}={N'}_{f}^{(2)}$ and $m_{a}^{(2)}={m'}_{a}^{(2)}$. In this way, we can regard $N_{f}^{(j)}={N'}_{f}^{(j)}$ and $m_{a}^{(j)}={m'}_{a}^{(j)}$, $j=1,2$ as node conditions for Bethe/Gauge correspondence. There is only one node for an ordinary Bethe/Gauge correspondence considered in \cite{DZ23}, and this point will be more clear for $A_r$ quiver gauge theory in the following of this paper.

\begin{equation}\label{}
\begin{aligned}
& \prod_{j\neq k}^{N_{2}}\dfrac{\text{sin}(\sigma_{j}^{(2)}\pm \sigma_{k}^{(2)}-m_{adj}^{(2)})}{\text{sin}(-\sigma_{j}^{(2)}\pm \sigma_{k}^{(2)}-m_{adj}^{(2)})}\prod_{a=1}^{N_{f}^{(2)}}\dfrac{\text{sin}(\sigma_{j}^{(2)}-m_{a}^{(2)})}{\text{sin}(-\sigma_{j}^{(2)}-m_{a}^{(2)})}\\
 &\times\prod_{k=1}^{N_{1}}\dfrac{\text{sin}(\sigma_{j}^{(2)}\pm \sigma_{k}^{(1)}-m_{\text{bfd}})}{\text{sin}(-\sigma_{j}^{(2)}\pm \sigma_{k}^{(1)}-m_{\text{bfd}})}\dfrac{\text{sin}^{2}(\sigma_{j}^{(2)}-m_{\text{adj}}^{(2)})}{\text{sin}^{2}(\sigma_{j}^{(2)}+m_{\text{adj}}^{(2)})}\dfrac{\text{sin}(\sigma_{j}^{(2)}-m_{\text{bfd}})}{\text{sin}(\sigma_{j}^{(2)}+m_{\text{bfd}})}=1
 \end{aligned}
\end{equation}
and
\begin{equation}\label{11}
\begin{aligned}
& \prod_{j\neq k}^{N_{2}}\dfrac{\text{sin}(\sigma_{j}^{(2)}\pm \sigma_{k}^{(2)}-m_{adj}^{(2)})}{\text{sin}(-\sigma_{j}^{(2)}\pm \sigma_{k}^{(2)}-m_{adj}^{(2)})}\prod_{a=1}^{N_{f}^{(2)}}\dfrac{\text{sin}(\sigma_{j}^{(2)}-m_{a}^{(2)})}{\text{sin}(-\sigma_{j}^{(2)}-m_{a}^{(2)})}\\
 &\times\prod_{k=1}^{N_{1}}\dfrac{\text{sin}(\sigma_{j}^{(2)}\pm \sigma_{k}^{(1)}-m_{\text{bfd}})}{\text{sin}(-\sigma_{j}^{(2)}\pm \sigma_{k}^{(1)}-m_{\text{bfd}})}\dfrac{\text{sin}^{2}(\sigma_{j}^{(2)}-m_{\text{adj}}^{(2)})}{\text{sin}^{2}(\sigma_{j}^{(2)}+m_{\text{adj}}^{(2)})}\dfrac{\text{sin}(\sigma_{j}^{(2)}-m_{\text{bfd}})}{\text{sin}(\sigma_{j}^{(2)}+m_{\text{bfd}})}=-1
 \end{aligned}
\end{equation}

We note that the Bethe ansatz equation (\ref{5}) and (\ref{6}) we want to map to is symmetric
about $ \sigma^{(1)} \leftrightarrow \sigma^{(2)}$ and $\xi_{\pm} \leftrightarrow \xi_{\pm}^{'}$. Using the vacuum equations (\ref{10}) and (\ref{11}), we choose 
\begin{equation}
\xi_{+}=-\dfrac{\eta}{2},\quad \xi_{-}=\dfrac{1}{2}
\end{equation}
To harmonize the above vacuum equations with the Bethe ansatz equations. Specifically, equation (\ref{10}) corresponds to equation (\ref{5}), and equation (\ref{11}) corresponds to equation (\ref{6}), respectively. This time we need $L_{1}-3=N_{f}^{(1)}$ and $L_{1}-4=N_{f}^{(2)}$. The parameter shifts here between $N_f-L$ are worthwhile to study its relationship with the duality of supersymmetric quiver gauge theories \cite{HKS11,YY15}.
And we need more constraints here
\begin{equation}
\theta_{1}=\theta_{2}=\dfrac{\eta}{2}, \quad \theta_{3}=0, \quad \theta_{4}=\dfrac{1}{2},\quad \theta_{5}=\theta_{6}=\dfrac{\eta}{2}, \quad \theta_{7}=0
\end{equation}
Here we use the dictionary
\begin{equation}\label{31}
\begin{aligned}
    &\pi \eta \longleftrightarrow m_{\text{adj}}^{(1)}=m_{\text{adj}}^{(2)},\quad \dfrac{\pi \eta}{2}\longleftrightarrow m_\text{bfd}\\
    &\{\pi\theta_{a}+\pi\dfrac{\eta}{2},-\pi\theta_{a}+\pi\frac{\eta}{2}\}\longleftrightarrow m_{a}^{(1)}=m_{a}^{(2)}
\end{aligned}
\end{equation}
Specially, we can set $m_{1}^{(1)}=m_{2}^{(1)}=\pi\eta$, $m_{4}^{(2)}=\pi\dfrac{\eta+1}{2}$, $m_{7}^{(2)}=\pi \dfrac{\eta}{2}$.

\subsection{$\text{Sp}(2N_1)\times \text{Sp}(2N_2)$}

For $\text{Sp}(2N_{1})\times \text{Sp}(2N_{2})$ quiver gauge theory, i.e. one gauge node (say the first node) is $\text{Sp}(2N_{1})$ gauge group and the other (the second node) is $\text{Sp}(2N_{2})$ gauge group, the bifundamental contribution to the effective potential is 
\begin{equation}
    \begin{aligned}
    W^{3d, \text{bfd}}_{\text{eff}}=&\dfrac{1}{\beta_{2}}\sum_{j=1}^{N_{1}}\sum_{k=1}^{N_{2}}\text{Li}_{2}(e^{-i(\pm\sigma_{j}^{(1)}\pm \sigma_{k}^{(2)})-im_{\text{bfd}}})-\dfrac{1}{4\beta_{2}}\sum_{j=1}^{N_{1}}\sum_{k=1}^{N_{2}}(\pm \sigma_{j}^{(1)}\pm \sigma_{k}^{(2)}+m_{\text{bfd}})^{2}\\
    &\dfrac{1}{\beta_{2}}\sum_{j=1}^{N_{2}}\sum_{k=1}^{N_{1}}\text{Li}_{2}(e^{-i(\pm\sigma_{j}^{(2)}\pm \sigma_{k}^{(1)})-im_{\text{bfd}}})-\dfrac{1}{4\beta_{2}}\sum_{j=1}^{N_{2}}\sum_{k=1}^{N_{1}}(\pm \sigma_{j}^{(2)}\pm \sigma_{k}^{(1)}+m_{\text{bfd}})^{2}
    \end{aligned}
\end{equation}
For the first node, the effective superpotential is 
\begin{equation*}\label{}
\begin{aligned}
W^{3d}_{\text{eff}}(\sigma,m)=&-\dfrac{2}{\beta_{2}}\sum_{j<k}^{N_{1}}\text{Li}_{2}(e^{i(\pm\sigma_{j}^{(1)}\pm \sigma_{k}^{(1)})})+\dfrac{1}{2\beta_{2}}\sum_{j<k}^{N_{1}}(\pm \sigma_{j}^{(1)}\pm \sigma_{k}^{(1)})^{2}\\
&+\dfrac{2}{\beta_{2}}\sum_{j<k}^{N_{1}}\text{Li}_{2}(e^{-i(\pm \sigma_{j}^{(1)}\pm \sigma_{k}^{(1)})-im_{adj}^{(1)}})-\dfrac{1}{2\beta_{2}}\sum_{j<k}^{N_{1}}(\pm \sigma_{j}^{(1)}\pm \sigma_{k}^{(1)}+m_{adj}^{(1)})^{2}\\
&+\dfrac{1}{\beta_{2}}\sum_{j=1}^{N_{1}}\sum_{a=1}^{N_{f}^{(1)}}\text{Li}_{2}(e^{-(\pm i \sigma_{j}^{(1)}+im_{a}^{(1)})})-\dfrac{1}{4\beta_{2}}\sum_{j=1}^{N_{1}}\sum_{a=1}^{N_{f}^{(1)}}(\pm \sigma_{j}^{(1)}+m_{a}^{(1)})^{2}\\
&-\dfrac{1}{\beta_{2}}\sum_{j=1}^{N_{1}}\text{Li}_{2}(e^{\pm 2i \sigma_{j}})+\dfrac{2}{\beta_{2}}\sigma_{j}^{2}\\
&+\dfrac{1}{\beta_{2}}\sum_{j=1}^{N_{1}}\text{Li}_{2}(e^{\pm 2i\sigma_{j}-im_{\text{adj}}^{(1)}})-\dfrac{1}{\beta_{2}}\sum_{j=1}^{N_{1}}(\sigma_{j}\pm m_{\text{adj}}^{(1)})^{2}\\
\end{aligned}
\end{equation*}
\begin{equation}
\begin{aligned}
\qquad \qquad \qquad &+\dfrac{1}{\beta_{2}}\sum_{j=1}^{N_{1}}\sum_{a=1}^{{N'}_{f}^{(1)}}\text{Li}_{2}(e^{\pm i \sigma_{j}^{(1)}-im_{a}^{(1)}})-\dfrac{1}{4\beta_{2}}\sum_{j=1}^{N_{1}}\sum_{a=1}^{{N'}_{f}^{(1)}}(\pm \sigma_{j}^{(1)}-m_{a}^{(1)})^{2}\\
&+\dfrac{1}{\beta_{2}}\sum_{j=1}^{N_{1}}\sum_{k=1}^{N_{2}}\text{Li}_{2}(e^{-i(\pm\sigma_{j}^{(1)}\pm \sigma_{k}^{(2)})-i m_{\text{bfd}}})-\dfrac{1}{4\beta_{2}}\sum_{j=1}^{N_{1}}\sum_{k=1}^{N_{2}}(\pm \sigma_{j}^{(1)}\pm \sigma_{k}^{(2)}+m_{\text{bfd}}))^{2}\\
&+\dfrac{1}{\beta_{2}}\sum_{j=1}^{N_{1}}\sum_{k=1}^{N_{2}}\text{Li}_{2}(e^{-i(\pm\sigma_{j}^{(1)}\pm \sigma_{k}^{(2)})-i m_{\text{bfd}}})-\dfrac{1}{4\beta_{2}}\sum_{j=1}^{N_{1}}\sum_{k=1}^{N_{2}}(\pm \sigma_{j}^{(1)}\pm \sigma_{k}^{(2)}+m_{\text{bfd}})^{2}
\end{aligned}
\end{equation}
The vacuum equation reads 
\begin{equation}\label{27}
\begin{aligned}
&\dfrac{\text{sin}^{2}(2\sigma_{j}^{(1)}-m_{\text{adj}}^{(1)})}{\text{sin}^{2}(2\sigma_{j}^{(1)}+m_{\text{adj}}^{(1)})}\prod_{j\neq k}^{N_{1}}\dfrac{\text{sin}^{2}(\sigma_{j}^{(1)}\pm \sigma_{k}^{(1)}-m_{adj}^{(1)})}{\text{sin}^{2}(-\sigma_{j}^{(1)}\pm \sigma_{k}^{(1)}-m_{adj}^{(1)})}\prod_{a=1}^{N_{f}^{(1)}}\dfrac{\text{sin}(\sigma_{j}^{(1)}-m_{a}^{(1)})}{\text{sin}(-\sigma_{j}^{(1)}-m_{a}^{(1)})}\\
&\times\prod_{a=1}^{{N'}_{f}^{(1)}}\dfrac{\text{sin}^{2}(\sigma_{j}^{(1)}-{m'}_{a}^{(1)})}{\text{sin}^{2}(-\sigma_{j}^{(1)}-{m'}_{a}^{(1)})}\prod_{k=1}^{N_{2}}\dfrac{\text{sin}^{2}(\sigma_{j}^{(1)}\pm \sigma_{k}^{(2)}-m_{\text{bfd}})}{\text{sin}^{2}(-\sigma_{j}^{(1)}\pm \sigma_{k}^{(2)}-m_{\text{bfd}})}=1
\end{aligned}
\end{equation}
In the case, $N_{f}^{(1)}={N'}_{f}^{(1)}$ and $m_{a}^{(1)}={m'}_{a}^{(1)}$, the square rooted vacuum equation of (\ref{27}) are
\begin{equation}\label{12}
    \begin{aligned}
        &\dfrac{\text{sin}(2\sigma_{j}^{(1)}-m_{\text{adj}}^{(1)})}{\text{sin}(2\sigma_{j}^{(1)}+m_{\text{adj}}^{(1)})}\prod_{j\neq k}^{N_{1}}\dfrac{\text{sin}(\sigma_{j}^{(1)}\pm \sigma_{k}^{(1)}-m_{adj}^{(1)})}{\text{sin}(-\sigma_{j}^{(1)}\pm \sigma_{k}^{(1)}-m_{adj}^{(1)})}\\
        &\times\prod_{a=1}^{N_{f}^{(1)}}\dfrac{\text{sin}(\sigma_{j}^{(1)}-m_{a}^{(1)})}{\text{sin}(-\sigma_{j}^{(1)}-m_{a}^{(1)})}\prod_{k=1}^{N_{2}}\dfrac{\text{sin}(\sigma_{j}^{(1)}\pm \sigma_{k}^{(2)}-m_{\text{bfd}})}{\text{sin}(-\sigma_{j}^{(1)}\pm \sigma_{k}^{(2)}-m_{\text{bfd}})}=1
    \end{aligned}
\end{equation}
and
\begin{equation}\label{}
    \begin{aligned}
        &\dfrac{\text{sin}(2\sigma_{j}^{(1)}-m_{\text{adj}}^{(1)})}{\text{sin}(2\sigma_{j}^{(1)}+m_{\text{adj}}^{(1)})}\prod_{j\neq k}^{N_{1}}\dfrac{\text{sin}(\sigma_{j}^{(1)}\pm \sigma_{k}^{(1)}-m_{adj}^{(1)})}{\text{sin}(-\sigma_{j}^{(1)}\pm \sigma_{k}^{(1)}-m_{adj}^{(1)})}\\
        &\times\prod_{a=1}^{N_{f}^{(1)}}\dfrac{\text{sin}(\sigma_{j}^{(1)}-m_{a}^{(1)})}{\text{sin}(-\sigma_{j}^{(1)}-m_{a}^{(1)})}\prod_{k=1}^{N_{2}}\dfrac{\text{sin}(\sigma_{j}^{(1)}\pm \sigma_{k}^{(2)}-m_{\text{bfd}})}{\text{sin}(-\sigma_{j}^{(1)}\pm \sigma_{k}^{(2)}-m_{\text{bfd}})}=-1
    \end{aligned}
\end{equation}
respectively.
For the second node, the effective superpotential is 
\begin{equation*}\label{}
\begin{aligned}
W^{3d}_{\text{eff}}(\sigma,m)=&-\dfrac{2}{\beta_{2}}\sum_{j<k}^{N_{2}}\text{Li}_{2}(e^{i(\pm\sigma_{j}^{(2)}\pm \sigma_{k}^{(2)})})+\dfrac{1}{2\beta_{2}}\sum_{j<k}^{N_{2}}(\pm \sigma_{j}^{(2)}\pm \sigma_{k}^{(2)})^{2}\\
&+\dfrac{2}{\beta_{2}}\sum_{j<k}^{N_{2}}\text{Li}_{2}(e^{-i(\pm \sigma_{j}^{(2)}\pm \sigma_{k}^{(2)})-im_{adj}^{(2)}})-\dfrac{1}{2\beta_{2}}\sum_{j<k}^{N_{2}}(\pm \sigma_{j}^{(2)}\pm \sigma_{k}^{(2)}+m_{adj}^{(2)})^{2}\\
\end{aligned}
\end{equation*}
\begin{equation}
\begin{aligned}
\qquad \qquad \qquad &+\dfrac{1}{\beta_{2}}\sum_{j=1}^{N_{2}}\sum_{a=1}^{N_{f}^{(2)}}\text{Li}_{2}(e^{-(\pm i \sigma_{j}^{(2)}+im_{a}^{(2)})})-\dfrac{1}{4\beta_{2}}\sum_{j=1}^{N_{2}}\sum_{a=1}^{N_{f}^{(2)}}(\pm \sigma_{j}^{(2)}+m_{a}^{(2)})^{2}\\
&-\dfrac{1}{\beta_{2}}\sum_{j=1}^{N_{2}}\text{Li}_{2}(e^{\pm 2i \sigma_{j}})+\dfrac{2}{\beta_{2}}\sigma_{2}^{2}\\
&+\dfrac{1}{\beta_{2}}\sum_{j=1}^{N_{2}}\text{Li}_{2}(e^{\pm 2i\sigma_{j}-im_{\text{adj}}^{(2)}})-\dfrac{1}{\beta_{2}}\sum_{j=1}^{N_{1}}(\sigma_{j}\pm m_{\text{adj}}^{(2)})^{2}\\
&+\dfrac{1}{\beta_{2}}\sum_{j=1}^{N_{2}}\sum_{a=1}^{{N'}_{f}^{(2)}}\text{Li}_{2}(e^{\pm i \sigma_{j}^{(2)}-im_{a}^{(2)}})-\dfrac{1}{4\beta_{2}}\sum_{j=1}^{N_{2}}\sum_{a=1}^{{N'}_{f}^{(2)}}(\pm \sigma_{j}^{(2)}-m_{a}^{(2)})^{2}\\
&+\dfrac{1}{\beta_{2}}\sum_{j=1}^{N_{1}}\sum_{k=1}^{N_{2}}\text{Li}_{2}(e^{-i(\pm\sigma_{j}^{(1)}\pm \sigma_{k}^{(2)})-i m_{\text{bfd}}})-\dfrac{1}{4\beta_{2}}\sum_{j=1}^{N_{1}}\sum_{k=1}^{N_{2}}(\pm \sigma_{j}^{(1)}\pm \sigma_{k}^{(2)}+m_{\text{bfd}}))^{2}\\
&+\dfrac{1}{\beta_{2}}\sum_{j=1}^{N_{1}}\sum_{k=1}^{N_{2}}\text{Li}_{2}(e^{-i(\pm\sigma_{j}^{(1)}\pm \sigma_{k}^{(2)})-i m_{\text{bfd}}})-\dfrac{1}{4\beta_{2}}\sum_{j=1}^{N_{1}}\sum_{k=1}^{N_{2}}(\pm \sigma_{j}^{(1)}\pm \sigma_{k}^{(2)}+m_{\text{bfd}})^{2}
\end{aligned}
\end{equation}
We write the vacuum equation as
\begin{equation}\label{28}
\begin{aligned}
&\dfrac{\text{sin}^{2}(2\sigma_{j}^{(2)}-m_{\text{adj}}^{(2)})}{\text{sin}^{2}(2\sigma_{j}^{(2)}+m_{\text{adj}}^{(2)})}\prod_{j\neq k}^{N_{2}}\dfrac{\text{sin}^{2}(\sigma_{j}^{(2)}\pm \sigma_{k}^{(2)}-m_{adj}^{(2)})}{\text{sin}^{2}(-\sigma_{j}^{(2)}\pm \sigma_{k}^{(2)}-m_{adj}^{(2)})}\prod_{a=1}^{N_{f}^{(2)}}\dfrac{\text{sin}(\sigma_{j}^{(2)}-m_{a}^{(2)})}{\text{sin}(-\sigma_{j}^{(2)}-m_{a}^{(2)})}\\
&\times\prod_{a=1}^{{N'}_{f}^{(2)}}\dfrac{\text{sin}(\sigma_{j}^{(2)}-{m'}_{a}^{(2)})}{\text{sin}(-\sigma_{j}^{(2)}-{m'}_{a}^{(2)})}\prod_{k=1}^{N_{1}}\dfrac{\text{sin}^{2}(\sigma_{j}^{(2)}\pm \sigma_{k}^{(1)}-m_{\text{bfd}})}{\text{sin}^{2}(-\sigma_{j}^{(2)}\pm \sigma_{k}^{(1)}-m_{\text{bfd}})}=1
\end{aligned}
\end{equation}
In the case $N_{f}^{(2)}={N'}_{f}^{(2)}$ and $m_{a}^{(2)}={m'}_{a}^{(2)}$, the vacuum equation is changed to the following two modalities of vacuum equations
\begin{equation}\label{13}
    \begin{aligned}
        &\dfrac{\text{sin}(2\sigma_{j}^{(2)}-m_{\text{adj}}^{(2)})}{\text{sin}(2\sigma_{j}^{(2)}+m_{\text{adj}}^{(2)})}\prod_{j\neq k}^{N_{2}}\dfrac{\text{sin}(\sigma_{j}^{(2)}\pm \sigma_{k}^{(2)}-m_{adj}^{(2)})}{\text{sin}(-\sigma_{j}^{(2)}\pm \sigma_{k}^{(2)}-m_{adj}^{(2)})}\\
        &\times\prod_{a=1}^{N_{f}^{(2)}}\dfrac{\text{sin}(\sigma_{j}^{(2)}-m_{a}^{(2)})}{\text{sin}(-\sigma_{j}^{(2)}-m_{a}^{(2)})}\prod_{k=1}^{N_{1}}\dfrac{\text{sin}(\sigma_{j}^{(2)}\pm \sigma_{k}^{(1)}-m_{\text{bfd}})}{\text{sin}(-\sigma_{j}^{(2)}\pm \sigma_{k}^{(1)}-m_{\text{bfd}})}=1
    \end{aligned}
\end{equation}
and
\begin{equation}
    \begin{aligned}
        &\dfrac{\text{sin}(2\sigma_{j}^{(2)}-m_{\text{adj}}^{(2)})}{\text{sin}(2\sigma_{j}^{(2)}+m_{\text{adj}}^{(2)})}\prod_{j\neq k}^{N_{2}}\dfrac{\text{sin}(\sigma_{j}^{(2)}\pm \sigma_{k}^{(2)}-m_{adj}^{(2)})}{\text{sin}(-\sigma_{j}^{(2)}\pm \sigma_{k}^{(2)}-m_{adj}^{(2)})}\\
        &\times \prod_{a=1}^{N_{f}^{(2)}}\dfrac{\text{sin}(\sigma_{j}^{(2)}-m_{a}^{(2)})}{\text{sin}(-\sigma_{j}^{(2)}-m_{a}^{(2)})}\prod_{k=1}^{N_{1}}\dfrac{\text{sin}(\sigma_{j}^{(2)}\pm \sigma_{k}^{(1)}-m_{\text{bfd}})}{\text{sin}(-\sigma_{j}^{(2)}\pm \sigma_{k}^{(1)}-m_{\text{bfd}})}=-1
    \end{aligned}
\end{equation}
if we take the square root of (\ref{28}). Using the symmetric and comparing with (\ref{5}) and (\ref{6}), we choose 
\begin{equation}
\xi_{+}=\dfrac{\eta}{2},\quad \xi_{-}=0
\end{equation}
to  reproduce the above vacuum equations from the Bethe ansatz equations. Specifically, equation (\ref{12}) corresponds to equation (\ref{5}) and equation (\ref{13}) corresponds to equation (\ref{6}). With the same dictionary (\ref{31}), we need to further augment two sites with $\theta_{1}=0, \theta_{2}=\dfrac{1}{2}$ and $L_{1}=N_{f}^{(1)}=N_{f}^{(2)}+2$.

\subsection{$\text{SO}(2N_1)\times \text{SO}(2N_2)$}

For $\text{SO}(2N_{1})\times \text{SO}(2N_{2})$ quiver gauge theory, i.e. one gauge node (say the first node) is $\text{SO}(2N_{1})$ gauge group and the other (the second node) is $\text{SO}(2N_{2})$ gauge group, the effective potential of the bifundamental multiplets reads
\begin{equation}
    \begin{aligned}
    W^{3d, \text{bfd}}_{\text{eff}}=&\dfrac{1}{\beta_{2}}\sum_{j=1}^{N_{1}}\sum_{k=1}^{N_{2}}\text{Li}_{2}(e^{-i(\pm\sigma_{j}^{(1)}\pm \sigma_{k}^{(2)})-im_{\text{bfd}}})-\dfrac{1}{4\beta_{2}}\sum_{j=1}^{N_{1}}\sum_{k=1}^{N_{2}}(\pm \sigma_{j}^{(1)}\pm \sigma_{k}^{(2)}+m_{\text{bfd}})^{2}\\
    &\dfrac{1}{\beta_{2}}\sum_{j=1}^{N_{2}}\sum_{k=1}^{N_{1}}\text{Li}_{2}(e^{-i(\pm\sigma_{j}^{(2)}\pm \sigma_{k}^{(1)})-im_{\text{bfd}}})-\dfrac{1}{4\beta_{2}}\sum_{j=1}^{N_{2}}\sum_{k=1}^{N_{1}}(\pm \sigma_{j}^{(2)}\pm \sigma_{k}^{(1)}+m_{\text{bfd}})^{2}
    \end{aligned}
\end{equation}
For the first node, the effective superpotential is 
\begin{equation}\label{}
\begin{aligned}
W^{3d}_{\text{eff}}(\sigma,m)=&-\dfrac{2}{\beta_{2}}\sum_{j<k}^{N_{1}}\text{Li}_{2}(e^{i(\pm\sigma_{j}^{(1)}\pm \sigma_{k}^{(1)})})+\dfrac{1}{2\beta_{2}}\sum_{j<k}^{N_{1}}(\pm \sigma_{j}^{(1)}\pm \sigma_{k}^{(1)})^{2}\\
&+\dfrac{2}{\beta_{2}}\sum_{j<k}^{N_{1}}\text{Li}_{2}(e^{-i(\pm \sigma_{j}^{(1)}\pm \sigma_{k}^{(1)})-im_{adj}^{(1)}})-\dfrac{1}{2\beta_{2}}\sum_{j<k}^{N_{1}}(\pm \sigma_{j}^{(1)}\pm \sigma_{k}^{(1)}+m_{adj}^{(1)})^{2}\\
&+\dfrac{1}{\beta_{2}}\sum_{j=1}^{N_{1}}\sum_{a=1}^{N_{f}^{(1)}}\text{Li}_{2}(e^{-(\pm i \sigma_{j}^{(1)}+im_{a}^{(1)})})-\dfrac{1}{4\beta_{2}}\sum_{j=1}^{N_{1}}\sum_{a=1}^{N_{f}^{(1)}}(\pm \sigma_{j}^{(1)}+m_{a}^{(1)})^{2}\\
&+\dfrac{1}{\beta_{2}}\sum_{j=1}^{N_{1}}\sum_{a=1}^{{N'}_{f}^{(1)}}\text{Li}_{2}(e^{\pm i \sigma_{j}^{(1)}-im_{a}^{(1)}})-\dfrac{1}{4\beta_{2}}\sum_{j=1}^{N_{1}}\sum_{a=1}^{{N'}_{f}^{(1)}}(\pm \sigma_{j}^{(1)}-m_{a}^{(1)})^{2}\\
&+\dfrac{1}{\beta_{2}}\sum_{j=1}^{N_{1}}\sum_{k=1}^{N_{2}}\text{Li}_{2}(e^{-i(\pm\sigma_{j}^{(1)}\pm \sigma_{k}^{(2)})-i m_{\text{bfd}}})-\dfrac{1}{4\beta_{2}}\sum_{j=1}^{N_{1}}\sum_{k=1}^{N_{2}}(\pm \sigma_{j}^{(1)}\pm \sigma_{k}^{(2)}+m_{\text{bfd}}))^{2}\\
&+\dfrac{1}{\beta_{2}}\sum_{j=1}^{N_{1}}\sum_{k=1}^{N_{2}}\text{Li}_{2}(e^{-i(\pm\sigma_{j}^{(1)}\pm \sigma_{k}^{(2)})-i m_{\text{bfd}}})-\dfrac{1}{4\beta_{2}}\sum_{j=1}^{N_{1}}\sum_{k=1}^{N_{2}}(\pm \sigma_{j}^{(1)}\pm \sigma_{k}^{(2)}+m_{\text{bfd}})^{2}
\end{aligned}
\end{equation}
The vacuum equation is found to be
\begin{equation}\label{29}
\begin{aligned}
&\prod_{j\neq k}^{N_{1}}\dfrac{\text{sin}^{2}(\sigma_{j}^{(1)}\pm \sigma_{k}^{(1)}-m_{adj}^{(1)})}{\text{sin}^{2}(-\sigma_{j}^{(1)}\pm \sigma_{k}^{(1)}-m_{adj}^{(1)})}\prod_{a=1}^{N_{f}^{(1)}}\dfrac{\text{sin}(\sigma_{j}^{(1)}-m_{a}^{(1)})}{\text{sin}(-\sigma_{j}^{(1)}-m_{a}^{(1)})}\\
&\times\prod_{a=1}^{{N'}_{f}^{(1)}}\dfrac{\text{sin}(\sigma_{j}^{(1)}-{m'}_{a}^{(1)})}{\text{sin}(-\sigma_{j}^{(1)}-{m'}_{a}^{(1)})}\prod_{k=1}^{N_{2}}\dfrac{\text{sin}^{2}(\sigma_{j}^{(1)}\pm \sigma_{k}^{(2)}-m_{\text{bfd}})}{\text{sin}^{2}(-\sigma_{j}^{(1)}\pm \sigma_{k}^{(2)}-m_{\text{bfd}})}=1
\end{aligned}
\end{equation}
With the square root of (\ref{29}), the vacuum equations are equivalently written as two forms according to the assumption $N_{f}^{(1)}={N'}_{f}^{(1)}$ and $m_{a}^{(1)}={m'}_{a}^{(1)}$
\begin{equation}\label{14}
    \prod_{j\neq k}^{N_{1}}\dfrac{\text{sin}(\sigma_{j}^{(1)}\pm \sigma_{k}^{(1)}-m_{adj}^{(1)})}{\text{sin}(-\sigma_{j}^{(1)}\pm \sigma_{k}^{(1)}-m_{adj}^{(1)})}\prod_{a=1}^{N_{f}^{(1)}}\dfrac{\text{sin}(\sigma_{j}^{(1)}-m_{a}^{(1)})}{\text{sin}(-\sigma_{j}^{(1)}-m_{a}^{(1)})}\prod_{k=1}^{N_{2}}\dfrac{\text{sin}(\sigma_{j}^{(1)}\pm \sigma_{k}^{(2)}-m_{\text{bfd}})}{\text{sin}(-\sigma_{j}^{(1)}\pm \sigma_{k}^{(2)}-m_{\text{bfd}})}=1
\end{equation}
and
\begin{equation}
    \prod_{j\neq k}^{N_{1}}\dfrac{\text{sin}(\sigma_{j}^{(1)}\pm \sigma_{k}^{(1)}-m_{adj}^{(1)})}{\text{sin}(-\sigma_{j}^{(1)}\pm \sigma_{k}^{(1)}-m_{adj}^{(1)})}\prod_{a=1}^{N_{f}^{(1)}}\dfrac{\text{sin}(\sigma_{j}^{(1)}-m_{a}^{(1)})}{\text{sin}(-\sigma_{j}^{(1)}-m_{a}^{(1)})}\prod_{k=1}^{N_{2}}\dfrac{\text{sin}(\sigma_{j}^{(1)}\pm \sigma_{k}^{(2)}-m_{\text{bfd}})}{\text{sin}(-\sigma_{j}^{(1)}\pm \sigma_{k}^{(2)}-m_{\text{bfd}})}=-1
\end{equation}
For the second node, the effective superpotemtial is 
\begin{equation}\label{}
\begin{aligned}
W^{3d}_{\text{eff}}(\sigma,m)=&-\dfrac{2}{\beta_{2}}\sum_{j<k}^{N_{2}}\text{Li}_{2}(e^{i(\pm\sigma_{j}^{(2)}\pm \sigma_{k}^{(2)})})+\dfrac{1}{2\beta_{2}}\sum_{j<k}^{N_{2}}(\pm \sigma_{j}^{(2)}\pm \sigma_{k}^{(2)})^{2}\\
&+\dfrac{2}{\beta_{2}}\sum_{j<k}^{N_{2}}\text{Li}_{2}(e^{-i(\pm \sigma_{j}^{(2)}\pm \sigma_{k}^{(2)})-im_{adj}^{(2)}})-\dfrac{1}{2\beta_{2}}\sum_{j<k}^{N_{2}}(\pm \sigma_{j}^{(2)}\pm \sigma_{k}^{(2)}+m_{adj}^{(2)})^{2}\\
&+\dfrac{1}{\beta_{2}}\sum_{j=1}^{N_{2}}\sum_{a=1}^{N_{f}^{(2)}}\text{Li}_{2}(e^{-(\pm i \sigma_{j}^{(2)}+im_{a}^{(2)})})-\dfrac{1}{4\beta_{2}}\sum_{j=1}^{N_{2}}\sum_{a=1}^{N_{f}^{(2)}}(\pm \sigma_{j}^{(2)}+m_{a}^{(2)})^{2}\\
&+\dfrac{1}{\beta_{2}}\sum_{j=1}^{N_{2}}\sum_{a=1}^{{N'}_{f}^{(2)}}\text{Li}_{2}(e^{\pm i \sigma_{j}^{(2)}-im_{a}^{(2)})})-\dfrac{1}{4\beta_{2}}\sum_{j=1}^{N_{2}}\sum_{a=1}^{{N'}_{f}^{(2)}}(\pm \sigma_{j}^{(2)}-m_{a}^{(2)})^{2}\\
&+\dfrac{1}{\beta_{2}}\sum_{j=1}^{N_{1}}\sum_{k=1}^{N_{2}}\text{Li}_{2}(e^{-i(\pm\sigma_{j}^{(1)}\pm \sigma_{k}^{(2)})-i m_{\text{bfd}}})-\dfrac{1}{4\beta_{2}}\sum_{j=1}^{N_{1}}\sum_{k=1}^{N_{2}}(\pm \sigma_{j}^{(1)}\pm \sigma_{k}^{(2)}+m_{\text{bfd}}))^{2}\\
&+\dfrac{1}{\beta_{2}}\sum_{j=1}^{N_{1}}\sum_{k=1}^{N_{2}}\text{Li}_{2}(e^{-i(\pm\sigma_{j}^{(1)}\pm \sigma_{k}^{(2)})-i m_{\text{bfd}}})-\dfrac{1}{4\beta_{2}}\sum_{j=1}^{N_{1}}\sum_{k=1}^{N_{2}}(\pm \sigma_{j}^{(1)}\pm \sigma_{k}^{(2)}+m_{\text{bfd}})^{2}
\end{aligned}
\end{equation}
The vacuum equation is written as
\begin{equation}\label{30}
\begin{aligned}
&\prod_{j\neq k}^{N_{2}}\dfrac{\text{sin}^{2}(\sigma_{j}^{(2)}\pm \sigma_{k}^{(2)}-m_{adj}^{(2)})}{\text{sin}^{2}(-\sigma_{j}^{(2)}\pm \sigma_{k}^{(2)}-m_{adj}^{(2)})}\prod_{a=1}^{N_{f}^{(2)}}\dfrac{\text{sin}(\sigma_{j}^{(2)}-m_{a}^{(2)})}{\text{sin}(-\sigma_{j}^{(2)}-m_{a}^{(2)})}\\
&\times\prod_{a=1}^{{N'}_{f}^{(2)}}\dfrac{\text{sin}(\sigma_{j}^{(2)}-{m'}_{a}^{(2)})}{\text{sin}(-\sigma_{j}^{(2)}-{m'}_{a}^{(2)})}\prod_{k=1}^{N_{1}}\dfrac{\text{sin}^{2}(\sigma_{j}^{(2)}\pm \sigma_{k}^{(1)}-m_{\text{bfd}})}{\text{sin}^{2}(-\sigma_{j}^{(2)}\pm \sigma_{k}^{(1)}-m_{\text{bfd}})}=1
\end{aligned}
\end{equation}
With the assumption $N_{f}^{(2)}={N'}_{f}^{(2)}$ and $m_{a}^{(2)}={m'}_{a}^{(2)}$, we obtain the following vacuum equations after we take the square root of (\ref{30})
\begin{equation}\label{}
    \prod_{j\neq k}^{N_{2}}\dfrac{\text{sin}(\sigma_{j}^{(2)}\pm \sigma_{k}^{(2)}-m_{adj}^{(2)})}{\text{sin}(-\sigma_{j}^{(2)}\pm \sigma_{k}^{(2)}-m_{adj}^{(2)})}\prod_{a=1}^{N_{f}^{(2)}}\dfrac{\text{sin}(\sigma_{j}^{(2)}-m_{a}^{(2)})}{\text{sin}(-\sigma_{j}^{(2)}-m_{a}^{(2)})}\prod_{k=1}^{N_{1}}\dfrac{\text{sin}(\sigma_{j}^{(2)}\pm \sigma_{k}^{(1)}-m_{\text{bfd}})}{\text{sin}(-\sigma_{j}^{(2)}\pm \sigma_{k}^{(1)}-m_{\text{bfd}})}=1
\end{equation}
and
\begin{equation}\label{15}
    \prod_{j\neq k}^{N_{2}}\dfrac{\text{sin}(\sigma_{j}^{(2)}\pm \sigma_{k}^{(2)}-m_{adj}^{(2)})}{\text{sin}(-\sigma_{j}^{(2)}\pm \sigma_{k}^{(2)}-m_{adj}^{(2)})}\prod_{a=1}^{N_{f}^{(2)}}\dfrac{\text{sin}(\sigma_{j}^{(2)}-m_{a}^{(2)})}{\text{sin}(-\sigma_{j}^{(2)}-m_{a}^{(2)})}\prod_{k=1}^{N_{1}}\dfrac{\text{sin}(\sigma_{j}^{(2)}\pm \sigma_{k}^{(1)}-m_{\text{bfd}})}{\text{sin}(-\sigma_{j}^{(2)}\pm \sigma_{k}^{(1)}-m_{\text{bfd}})}=-1
\end{equation}
respectively. With the same map as (\ref{31}), we choose 
\begin{equation}
\xi_{+}=-\dfrac{\eta}{2},\quad \xi_{-}=\dfrac{1}{2}
\end{equation}
to fit the above vacuum equations with the Bethe ansatz equations. Specifically, equation (\ref{14}) corresponds to equation (\ref{5}) and equation (\ref{15}) corresponds to equation (\ref{6}). In this way, we need to further append two sites with 
\begin{equation}
\theta_{1}=\dfrac{1}{2}, \quad \theta_{2}=0
\end{equation}
and $L_{1}=N_{f}^{(1)}={N'}_{f}^{(2)}+2$. We notice that the $m_{1}^{(2)}=\dfrac{\pi}{2}$ and $m_{2}^{(2)}=\pi \dfrac{1+\eta}{2}$.

\subsection{$\text{SO}(2N_1)\times \text{SO}(2N_2+1)$}

For $\text{SO}(2N_{1})\times \text{SO}(2N_{2}+1)$ quiver gauge theory, i.e. one gauge 
node (say the first node) is $\text{SO}(N_{1})$ gauge group and the other (the second node) is $\text{SO}(2N_{2}+1)$ gauge group, the bifundamental contribution to the effective potential is
\begin{equation}
    \begin{aligned}
    W^{3d, \text{bdf}}_{\text{eff}}=&\dfrac{1}{\beta_{2}}\sum_{j=1}^{N_{1}}\sum_{k=1}^{N_{2}}\text{Li}_{2}(e^{-i(\pm\sigma_{j}^{(1)}\pm \sigma_{k}^{(2)})-im_{\text{bfd}}})-\dfrac{1}{4\beta_{2}}\sum_{j=1}^{N_{1}}\sum_{k=1}^{N_{2}}(\pm \sigma_{j}^{(1)}\pm \sigma_{k}^{(2)}+m_{\text{bfd}})^{2}\\
    &+\dfrac{1}{\beta_{2}}\sum_{j=1}^{N_{2}}\sum_{k=1}^{N_{1}}\text{Li}_{2}(e^{-i(\pm\sigma_{j}^{(2)}\pm \sigma_{k}^{(1)})-im_{\text{bfd}}})-\dfrac{1}{4\beta_{2}}\sum_{j=1}^{N_{2}}\sum_{k=1}^{N_{1}}(\pm \sigma_{j}^{(2)}\pm \sigma_{k}^{(1)}+m_{\text{bfd}})^{2}\\
    &+\dfrac{1}{\beta_{2}}\sum_{j=1}^{N_{1}}\text{Li}_{2}(e^{-i(\pm\sigma_{j}^{(1)})-im_{\text{bfd}}})-\dfrac{1}{4\beta_{2}}\sum_{j=1}^{N_{1}}(\pm\sigma_{j}^{(1)}+m_{\text{bfd}})^{2}\\
    &+\dfrac{1}{\beta_{2}}\sum_{j=1}^{N_{1}}\text{Li}_{2}(e^{-i(\pm\sigma_{j}^{(1)})-im_{\text{bfd}}})-\dfrac{1}{4\beta_{2}}\sum_{j=1}^{N_{1}}(\pm\sigma_{j}^{(1)}+m_{\text{bfd}})^{2}
    \end{aligned}
\end{equation}
For these nodes, the two vacuum equation are
\begin{equation}\label{}
\begin{aligned}
&\prod_{j\neq k}^{N_{1}}\dfrac{\text{sin}^{2}(\sigma_{j}^{(1)}\pm \sigma_{k}^{(1)}-m_{adj}^{(1)})}{\text{sin}^{2}(-\sigma_{j}^{(1)}\pm \sigma_{k}^{(1)}-m_{adj}^{(1)})}\prod_{a=1}^{N_{f}^{(1)}}\dfrac{\text{sin}(\sigma_{j}^{(1)}-m_{a}^{(1)})}{\text{sin}(-\sigma_{j}^{(1)}-m_{a}^{(1)})}\prod_{a=1}^{{N'}_{f}^{(1)}}\dfrac{\text{sin}(\sigma_{j}^{(1)}-{m'}_{a}^{(1)})}{\text{sin}(-\sigma_{j}^{(1)}-{m'}_{a}^{(1)})}\\
&\times\prod_{k=1}^{N_{2}}\dfrac{\text{sin}^{2}(\sigma_{j}^{(1)}\pm \sigma_{k}^{(2)}-m_{\text{bfd}})}{\text{sin}^{2}(-\sigma_{j}^{(1)}\pm \sigma_{k}^{(2)}-m_{\text{bfd}})}\dfrac{\text{sin}^{2}(\sigma_{j}^{(1)}-m_{\text{bfd}})}{\text{sin}^{2}(\sigma_{j}^{(1)}+m_{\text{bfd}})}=1
\end{aligned}
\end{equation}
and 
 \begin{equation}\label{55}
\begin{aligned}
&\dfrac{\text{sin}^{4}(\sigma_{j}^{(2)}-m_{\text{adj}}^{(2)})}{\text{sin}^{4}(\sigma_{j}^{(2)}+m_{\text{adj}}^{(2)})}\prod_{j\neq k}^{N_{2}}\dfrac{\text{sin}^{2}(\sigma_{j}^{(2)}\pm \sigma_{k}^{(2)}-m_{adj}^{(2)})}{\text{sin}^{2}(-\sigma_{j}^{(2)}\pm \sigma_{k}^{(2)}-m_{adj}^{(2)})}\prod_{a=1}^{N_{f}^{(2)}}\dfrac{\text{sin}(\sigma_{j}^{(2)}-m_{a}^{(2)})}{\text{sin}(-\sigma_{j}^{(2)}-m_{a}^{(2)})}\\
&\times\prod_{a=1}^{{N'}_{f}^{(2)}}\dfrac{\text{sin}(\sigma_{j}^{(2)}-{m'}_{a}^{(2)})}{\text{sin}(-\sigma_{j}^{(2)}-{m'}_{a}^{(2)})}\prod_{k=1}^{N_{1}}\dfrac{\text{sin}^{2}(\sigma_{j}^{(2)}\pm \sigma_{k}^{(1)}-m_{\text{bfd}})}{\text{sin}^{2}(-\sigma_{j}^{(2)}\pm \sigma_{k}^{(1)}-m_{\text{bfd}})}=1
\end{aligned}
\end{equation}
respectively. Following the same approach, we see that  
\begin{equation}
\xi_{-}=\dfrac{1}{2}, \quad  \xi_{+}=-\dfrac{\eta}{2}
\end{equation}
to match the Bethe equations (\ref{5}) and (\ref{6}), where we need to choose $\theta_{1}=0$, $\theta_{2}=\dfrac{1}{2}$, $\theta_{3}=\theta_{4}=\dfrac{\eta}{2}$ and $L_{1}=N_{f}^{(1)}+1=N_{f}^{(2)}+3$. 

\subsection{$\text{Sp}(2N_1)\times \text{SO}(2N_2)$}

For $\text{Sp}(2N_{1})\times \text{SO}(2N_{2})$ quiver gauge theory, i.e. one gauge 
node (say the first node) is $\text{Sp}(2N_{1})$ gauge group and the other (the second node) is $\text{SO}(2N_{2})$ gauge group, the bifundamental contribution to the effective potential is
\begin{equation}
    \begin{aligned}
    W^{3d, \text{bdf}}_{\text{eff}}=&\dfrac{1}{\beta_{2}}\sum_{j=1}^{N_{1}}\sum_{k=1}^{N_{2}}\text{Li}_{2}(e^{-i(\pm\sigma_{j}^{(1)}\pm \sigma_{k}^{(2)})-im_{\text{bfd}}})-\dfrac{1}{4\beta_{2}}\sum_{j=1}^{N_{1}}\sum_{k=1}^{N_{2}}(\pm \sigma_{j}^{(1)}\pm \sigma_{k}^{(2)}+m_{\text{bfd}})^{2}\\
    &+\dfrac{1}{\beta_{2}}\sum_{j=1}^{N_{2}}\sum_{k=1}^{N_{1}}\text{Li}_{2}(e^{-i(\pm\sigma_{j}^{(2)}\pm \sigma_{k}^{(1)})-im_{\text{bfd}}})-\dfrac{1}{4\beta_{2}}\sum_{j=1}^{N_{2}}\sum_{k=1}^{N_{1}}(\pm \sigma_{j}^{(2)}\pm \sigma_{k}^{(1)}+m_{\text{bfd}})^{2}
    \end{aligned}
\end{equation}

For the two nodes, the vacuum equation are (\ref{27}) and (\ref{30}), respectivly. So we just need to compare the vacuum equation (\ref{13}) with the Bethe ansatz equation (\ref{5}) and compare the vacuum equation (\ref{15}) with the Bethe ansatz equation (\ref{6}). Then we choose the boundary condition 
\begin{equation}
    \xi_{-}=0,\quad \xi_{+}=-\dfrac{\eta}{2}
\end{equation}
to recover the vacuum equation with Bethe ansatz equation. We need to add the condition $\theta_{1}=\dfrac{1}{2}$, $\theta_{2}=0$ and $L_{1}=N_{f}^{(1)}+2=N_{f}^{(2)}$. If we consider $\text{SO}(2N_{1})\times \text{Sp}(2N_{2})$ quiver gauge theory, the boundary condition is the same to the above. We choose $\theta_{1}=\dfrac{1}{2}$, $\theta_{2}=0$ and $L_{1}=N_{f}^{(1)}=N_{f}^{(2)}+2$.

\subsection{$\text{Sp}(2N_1)\times \text{SO}(2N_2+1)$}

For $\text{Sp}(2N_{1})\times \text{SO}(2N_{2}+1)$ quiver gauge theory, i.e. one gauge node (the first node) is $\text{Sp}(2N_{1})$ gauge group and the other (the second one) is $\text{SO}(2N_{2}+1)$ gauge group, the contribution of bifundamental matter to the effective potential is
\begin{equation}
    \begin{aligned}
    W^{3d, \text{bdf}}_{\text{eff}}=&\dfrac{1}{\beta_{2}}\sum_{j=1}^{N_{1}}\sum_{k=1}^{N_{2}}\text{Li}_{2}(e^{-i(\pm\sigma_{j}^{(1)}\pm \sigma_{k}^{(2)})-im_{\text{bfd}}})-\dfrac{1}{4\beta_{2}}\sum_{j=1}^{N_{1}}\sum_{k=1}^{N_{2}}(\pm \sigma_{j}^{(1)}\pm \sigma_{k}^{(2)}+m_{\text{bfd}})^{2}\\
    &+\dfrac{1}{\beta_{2}}\sum_{j=1}^{N_{2}}\sum_{k=1}^{N_{1}}\text{Li}_{2}(e^{-i(\pm\sigma_{j}^{(2)}\pm \sigma_{k}^{(1)})-im_{\text{bfd}}})-\dfrac{1}{4\beta_{2}}\sum_{j=1}^{N_{2}}\sum_{k=1}^{N_{1}}(\pm \sigma_{j}^{(2)}\pm \sigma_{k}^{(1)}+m_{\text{bfd}})^{2}\\
    &+\dfrac{1}{\beta_{2}}\sum_{j=1}^{N_{1}}\text{Li}_{2}(e^{-i(\pm\sigma_{j}^{(1)})-im_{\text{bfd}}})-\dfrac{1}{4\beta_{2}}\sum_{j=1}^{N_{1}}(\pm\sigma_{j}^{(1)}+m_{\text{bfd}})^{2}\\
    &+\dfrac{1}{\beta_{2}}\sum_{j=1}^{N_{1}}\text{Li}_{2}(e^{-i(\pm\sigma_{j}^{(1)})-im_{\text{bfd}}})-\dfrac{1}{4\beta_{2}}\sum_{j=1}^{N_{1}}(\pm\sigma_{j}^{(1)}+m_{\text{bfd}})^{2}
    \end{aligned}
\end{equation}
For the first node, the vacuum equation is
\begin{equation}\label{}
\begin{aligned}
&\dfrac{\text{sin}^{2}(2\sigma_{j}^{(1)}-m_{\text{adj}}^{(1)})}{\text{sin}^{2}(2\sigma_{j}^{(1)}+m_{\text{adj}}^{(1)})}\prod_{j\neq k}^{N_{1}}\dfrac{\text{sin}^{2}(\sigma_{j}^{(1)}\pm \sigma_{k}^{(1)}-m_{adj}^{(1)})}{\text{sin}^{2}(-\sigma_{j}^{(1)}\pm \sigma_{k}^{(1)}-m_{adj}^{(1)})}\prod_{a=1}^{N_{f}^{(1)}}\dfrac{\text{sin}(\sigma_{j}^{(1)}-m_{a}^{(1)})}{\text{sin}(-\sigma_{j}^{(1)}-m_{a}^{(1)})}\\
&\times\prod_{a=1}^{{N'}_{f}^{(1)}}\dfrac{\text{sin}^{2}(\sigma_{j}^{(1)}-{m'}_{a}^{(1)})}{\text{sin}^{2}(-\sigma_{j}^{(1)}-{m'}_{a}^{(1)})}\prod_{k=1}^{N_{2}}\dfrac{\text{sin}^{2}(\sigma_{j}^{(1)}\pm \sigma_{k}^{(2)}-m_{\text{bfd}})}{\text{sin}^{2}(-\sigma_{j}^{(1)}\pm \sigma_{k}^{(2)}-m_{\text{bfd}})}\dfrac{\text{sin}^{2}(\sigma_{j}^{(1)}-m_{\text{bfd}})}{\text{sin}^{2}(\sigma_{j}^{(1)}+m_{\text{bfd}})}=1
\end{aligned}
\end{equation}
For the second node, the vacuum equation is (\ref{55}). So we use the same method called square root to correspond to the Bethe equations (\ref{5}) and (\ref{6}). Then we get $L_{1}=N_{f}^{(1)}+1=N_{f}^{(2)}+2$, $\theta_{1}=0$, $\theta_{2}=\dfrac{1}{2}$ and $\theta_{3}=\dfrac{\eta}{2}$. The boundary condition is
\begin{equation}
    \xi_{-}=0,\quad \xi_{+}=\dfrac{\eta}{2}
\end{equation}
If we consider $\text{SO}(2N_{1}+1)\times \text{Sp}(2N_{2})$ quiver gauge theory, the boundary condition will be $\xi_{-}=0$, $\xi_{+}=-\dfrac{\eta}{2}$. And we choose $\theta_{1}=\theta_{2}=\dfrac{\eta}{2}$, $\theta_{3}=\theta_{4}=0$, $\theta_{5}=\dfrac{1}{2}$ and $L_{1}=N_{f}^{(1)}+2=N_{f}^{(2)}+3$.

It is interesting to note that, the size $L$ of the spin chain is corresponding to the rank of 
the flavor symmetry for the supersymmetric quiver gauge theory. With an appropriate 
chosen boundary condition, we can make sure that $N_f^{(1)}-N_f^{(2)}\geq 1$. In the 
other word, it makes sure that $\text{SU}(N_f^{(2)})$ is a subgroup of $\text{SU}(N_f^{(1)})$. 

\subsection{2d limit}
In 2d limit, $[x]\rightarrow x$, the first Bethe ansatz equation (\ref{5}) of $sl_{3}$ spin chain degenerates to 
\begin{equation}
\begin{aligned}
    &\dfrac{(2u_{i}^{(1)}-\eta)(u_{i}^{(1)}+\xi_{-}+\frac{\eta}{2})(u_{i}^{(1)}-\xi_{+})}{(2u_{i}^{(1)}+\eta)(u_{i}^{(1)}-\xi_{-}-\frac{\eta}{2})(u_{i}^{(1)}+\xi_{+})}\prod_{j=1}^{N_{1}}\dfrac{(u_{i}^{(1)}-u_{j}^{(1)}-\eta)(u_{i}^{(1)}+u_{j}^{(1)}-\eta)}{(u_{i}^{(1)}-u_{j}^{(1)}+\eta)(u_{i}^{(1)}+u_{j}^{(1)}+\eta)}\\
    &\times\prod_{k=1}^{N_{2}}\dfrac{(u_{i}^{(1)}-u_{k}^{(2)}-\frac{\eta}{2})(u_{i}^{(1)}+u_{k}^{(2)}-\frac{\eta}{2})}{(u_{i}^{(1)}-u_{k}^{(2)}+\frac{\eta}{2})(u_{i}^{(1)}+u_{k}^{(2)}+\frac{\eta}{2})}\prod_{a=1}^{L_{1}}\dfrac{(u_{i}^{(1)}+\theta_{a}-\frac{\eta}{2})(u_{i}^{(1)}-\theta_{a}-\frac{\eta}{2})}{(u_{i}^{(1)}+\theta_{a}+\frac{\eta}{2})(u_{i}^{(1)}-\theta_{a}+\frac{\eta}{2})}=1
\end{aligned}
\end{equation}
the second Bethe ansatz equation (\ref{6}) degenerates to
\begin{equation}
\begin{aligned}
    &\dfrac{(2u_{i}^{(2)}+\eta)(u_{i}^{(2)}+\xi_{-})(u_{i}^{(2)}-\xi_{+}-\frac{\eta}{2})}{(2u_{i}^{(2)}-\eta)(u_{i}^{(1)}-\xi_{-})(u_{i}^{(1)}+\xi_{+}+\frac{\eta}{2})}\prod_{j=1}^{N_{2}}\dfrac{(u_{i}^{(2)}-u_{j}^{(2)}+\eta)(u_{i}^{(2)}+u_{j}^{(2)}+\eta)}{(u_{i}^{(2)}-u_{j}^{(2)}-\eta)(u_{i}^{(2)}+u_{j}^{(2)}-\eta)}\\
    &\times\prod_{k=1}^{N_{1}}\dfrac{(u_{i}^{(2)}-u_{k}^{(1)}+\frac{\eta}{2})(u_{i}^{(2)}+u_{k}^{(1)}+\frac{\eta}{2})}{(u_{i}^{(2)}-u_{k}^{(1)}-\frac{\eta}{2})(u_{i}^{(2)}+u_{k}^{(1)}-\frac{\eta}{2})}=1
\end{aligned}
\end{equation}
For 2d $\text{Sp}(2N_{1})\times \text{SO}(2N_{1})$ quiver gauge theory, the first vacuum equation (\ref{27}) degenerates to
\begin{equation}\label{36}
\begin{aligned}
&\dfrac{(2\sigma_{j}^{(1)}-m_{\text{adj}}^{(1)})^{2}}{(2\sigma_{j}^{(2}+m_{\text{adj}}^{(1)})^{2}}\prod_{j\neq k}^{N_{1}}\dfrac{(\sigma_{j}^{(1)}\pm \sigma_{k}^{(1)}-m_{\text{adj}}^{(1)})^{2}}{(-\sigma_{j}^{(1)}\pm \sigma_{k}^{(1)}-m_{\text{adj}}^{(1)})^{2}}\prod_{a=1}^{N_{f}^{(1)}}\dfrac{(\sigma_{j}^{(1)}-m_{a}^{(1)})}{(-\sigma_{j}^{(1)}-m_{a}^{(1)})}\\
&\times\prod_{a=1}^{{N'}_{f}^{(1)}}\dfrac{(\sigma_{j}^{(1)}-{m'}_{a}^{(1)})}{(-\sigma_{j}^{(1)}-{m'}_{a}^{(1)})}\prod_{k=1}^{N_{2}}\dfrac{(\sigma_{j}^{(1)}\pm \sigma_{k}^{(2)}-m_{\text{bfd}})^{2}}{(-\sigma_{j}^{(1)}\pm \sigma_{k}^{(2)}-m_{\text{bfd}})^{2}}=1
\end{aligned}
\end{equation}
Taking the square root of (\ref{36}), we can write the vacuum equations equivalently with the identification $N_{f}^{(1)}={N'}_{f}^{(1)}$ and $m_{a}^{(1)}={m'}_{a}^{(1)}$
\begin{equation}\label{37}
\begin{aligned}
&\dfrac{(2\sigma_{j}^{(1)}-m_{\text{adj}}^{(1)})}{(2\sigma_{j}^{(2}+m_{\text{adj}}^{(1)})}\prod_{j\neq k}^{N_{1}}\dfrac{(\sigma_{j}^{(1)}\pm \sigma_{k}^{(1)}-m_{\text{adj}}^{(1)})}{(-\sigma_{j}^{(1)}\pm \sigma_{k}^{(1)}-m_{\text{adj}}^{(1)})}\prod_{a=1}^{N_{f}^{(1)}}\dfrac{(\sigma_{j}^{(1)}-m_{a}^{(1)})}{(-\sigma_{j}^{(1)}-m_{a}^{(1)})}\\
&\times \prod_{k=1}^{N_{2}}\dfrac{(\sigma_{j}^{(1)}\pm \sigma_{k}^{(2)}-m_{\text{bfd}})}{(-\sigma_{j}^{(1)}\pm \sigma_{k}^{(2)}-m_{\text{bfd}})}=1
\end{aligned}
\end{equation}
and
\begin{equation}\label{}
\begin{aligned}
&\dfrac{(2\sigma_{j}^{(1)}-m_{\text{adj}}^{(1)})}{(2\sigma_{j}^{(2}+m_{\text{adj}}^{(1)})}\prod_{j\neq k}^{N_{1}}\dfrac{(\sigma_{j}^{(1)}\pm \sigma_{k}^{(1)}-m_{\text{adj}}^{(1)})}{(-\sigma_{j}^{(1)}\pm \sigma_{k}^{(1)}-m_{\text{adj}}^{(1)})}\prod_{a=1}^{N_{f}^{(1)}}\dfrac{(\sigma_{j}^{(1)}-m_{a}^{(1)})}{(-\sigma_{j}^{(1)}-m_{a}^{(1)})}\\
&\times \prod_{k=1}^{N_{2}}\dfrac{(\sigma_{j}^{(1)}\pm \sigma_{k}^{(2)}-m_{\text{bfd}})}{(-\sigma_{j}^{(1)}\pm \sigma_{k}^{(2)}-m_{\text{bfd}})}=-1
\end{aligned}
\end{equation}
respectively. The second vacuum equation (\ref{30}) degenerates to is 
\begin{equation}\label{}
\begin{aligned}
&\prod_{j\neq k}^{N_{2}}\dfrac{(\sigma_{j}^{(2)}\pm \sigma_{k}^{(2)}-m_{\text{adj}}^{(2)})^{2}}{(-\sigma_{j}^{(2)}\pm \sigma_{k}^{(2)}-m_{\text{adj}}^{(2)})^{2}}\prod_{a=1}^{N_{f}^{(2)}}\dfrac{(\sigma_{j}^{(2)}-m_{a}^{(2)})}{(-\sigma_{j}^{(2)}-m_{a}^{(2)})}\\
&\times\prod_{a=1}^{{N'}_{f}^{(2)}}\dfrac{(\sigma_{j}^{(2)}-{m'}_{a}^{(2)})}{(-\sigma_{j}^{(2)}-{m'}_{a}^{(2)})}\prod_{k=1}^{N_{1}}\dfrac{(\sigma_{j}^{(2)}\pm \sigma_{k}^{(1)}-m_{\text{bfd}})^{2}}{(-\sigma_{j}^{(2)}\pm \sigma_{k}^{(1)}-m_{\text{bfd}})^{2}}=1
\end{aligned}
\end{equation}
In this way
\begin{equation}\label{35}
    \eta \longleftrightarrow m_{\text{adj}},\quad 
    \dfrac{\eta}{2} \longleftrightarrow m_{\text{bfd}},\quad \{\theta_{a}+\frac{\eta}{2},-\theta_{a}+\dfrac{\eta}{2}\} \longleftrightarrow m_{a}
\end{equation}
we choose 
\begin{equation}
    \xi_{-}=\xi_{+}=0
\end{equation}
to get the duality between the two vacuum equations and the two Bethe ansatz equations. We also need to add $\theta_{1}=0$ and $L_{1}=N_{f}^{(1)}=N_{f}^{(2)}$ in the correspondence. In the first vacuum equation, we can notice $m_{1}^{(1)}=\dfrac{\eta}{2}$. If we let extra $\theta_{2}=0$ and ${m'}_{2}^{(1)}=\dfrac{\eta}{2}$, the first vacuum equation (\ref{37}) is the same to the first vacuum equation of 2d $\text{Sp}(2N_{1})\times \text{SO}(2N_{1})$ quiver gauge theory in \cite{KZ21}. At this time we have $L_{1}-1=N_{f}^{(1)}$. 

Using the same method as 2d $\text{Sp}(N_{1})\times \text{SO}(2N_{1})$ quiver gauge theory, we can get the results of other gauge groups. For 2d $\text{Sp}(2N_{1})\times \text{SO}(2N_{2}+1)$ quiver gauge theory, we just write the results 
\begin{equation}
\xi_{-}=\xi_{+}=0
\end{equation}
to match the vacuum equations with the Bethe ansatz equations. We need to further augment $\theta_{1}=0$, $\theta_{2}=\theta_{3}=\dfrac{\eta}{2}$ and $L_{1}=N_{f}^{(1)}+1=N_{f}^{(2)}+2$ with the map (\ref{35}). If we let extra parameter $\theta_{4}=0$ and $m_{4^{'}}^{(2)}=\dfrac{\eta}{2}$, the vacuum equation (\ref{37}) is the same to the vacuum equation 2d $\text{Sp}(2N_{1})\times \text{SO}(2N_{2}+1)$ quiver gauge theory in \cite{KZ21}. And $L_{1}-2=N_{f}^{(1)}$.

From the values of parameters above, we can see the vacuum equation of $A_{2}$ quiver gauge theory in \cite{KZ21} is a special case of our results.

For 2d $\text{SO}(2N_{1}+1)\times \text{SO}(2N_{2}+1)$ quiver gauge theories, we have the same boundary 
\begin{equation}
    \xi_{-}=\xi_{+}=0
\end{equation}
to fit the vacuum equations with the Bethe ansatz equations. We need to further append $\theta_{1}=\theta_{2}=\theta_{3}=\theta_{4}=\dfrac{\eta}{2}$ and $L_{1}-2=N_{f}^{(1)}=N_{f}^{(2)}$ with the dictionary (\ref{35}).

For 2d $\text{Sp}(N_{1})\times \text{Sp}(N_{2})$ quiver gauge theories, we have the same boundary condition
\begin{equation}
    \xi_{-}=0,\quad \xi_{+}=\dfrac{\eta}{2}
\end{equation}
to equal the vacuum equations with the Bethe ansatz equations. We need to further add $\theta_{1}=\dfrac{\eta}{2}$ and $L_{1}=N_{f}^{(1)}=N_{f}^{(2)}+1$ in this relation (\ref{35}).

For 2d $\text{SO}(2N_{1})\times \text{SO}(2N_{2})$ quiver gauge theories, we have the same boundary condition
\begin{equation}
    \xi_{-}=\xi_{+}=0
\end{equation}
to harmonize the vacuum equations with the Bethe ansatz equations. We need to further assume  $L_{1}=N_{f}^{(1)}=N_{f}^{(2)}$ under the map (\ref{35}).

For 2d $\text{SO}(2N_{1})\times \text{SO}(2N_{2}+1)$ quiver gauge theories, we choose the boundary condition
\begin{equation}
\xi_{-}=0, \quad \xi_{+}=\dfrac{\eta}{2}
\end{equation}
to match the vacuum equations with the Bethe equations, where we let $\theta_{1}=0$, $\theta_{2}=\dfrac{\eta}{2}$ and $L_{1}=N_{f}^{(2)}+2$.

\section{$A_{r}$ quiver gauge theory}\label{d}

For 3d $A_{r}$ quiver gauge theory with product gauge group  $G=SU(N_{1})\times \cdots \times SU(N_{r})$, we consider the vacuum equation in each node with our new effective superpotential. For the node $\mathbf{i}$, $\mathbf{i}=1,\cdots, r$, the vacuum equation can be written as
\begin{equation}\label{}
\begin{aligned}
&\prod_{k\neq j}^{N_{\mathbf{i}}}\dfrac{\text{sin}^{2}(\sigma_{j}^{(\mathbf{i})}-\sigma_{k}^{(\mathbf{i})}+m_{\text{adj}}^{(\mathbf{i})})}{\text{sin}^{2}(\sigma_{j}^{(\mathbf{i})}-\sigma_{k}^{(\mathbf{i})}-m_{\text{adj}}^{(\mathbf{i})})}
\prod_{l=1}^{N_{\mathbf{i}+1}}\dfrac{\text{sin}^{2}(\sigma_{l}^{(\mathbf{i})}-\sigma_{j}^{(\mathbf{i}+1)}-m_{\text{bfd}}^{(\mathbf{i,i+1})})}{\text{sin}^{2}(\sigma_{l}^{(\mathbf{i})}-\sigma_{j}^{(\mathbf{i}+1)}+m_{\text{bfd}}^{(\mathbf{i,i+1})})}\\
&\times \prod_{l=1}^{N_{\mathbf{i}-1}}\dfrac{\text{sin}^{2}(\sigma_{l}^{(\mathbf{i})}-\sigma_{j}^{(\mathbf{i}-1)}-m_{\text{bfd}}^{(\mathbf{i-1,i})})}{\text{sin}^{2}(\sigma_{l}^{(\mathbf{i})}-\sigma_{j}^{(\mathbf{i}-1)}+m_{\text{bfd}}^{(\mathbf{i-1,i})})}=\prod_{a=1}^{N_{f}^{(\mathbf{i})}}\dfrac{\text{sin}^{2}(\sigma_{j}^{(\mathbf{i})}+m_{a}^{(\mathbf{i})})}{\text{sin}^{2}(\sigma_{j}^{(\mathbf{i})}-m_{a}^{'(\mathbf{i})})}
\end{aligned}
\end{equation}
where $N_{f}^{(\mathbf{i})}$ is the dimension of the fundamental representation in the $\mathbf{i}$-th gauge node. Taking the 2d limit, we can get the vacuum equation of 2d quiver gauge theory with product gauge group  $G=SU(N_{1})\times \cdots \times SU(N_{r})$, 
\begin{equation}\label{}
\begin{aligned}
&\prod_{k\neq j}^{N_{\mathbf{i}}}\dfrac{(\sigma_{j}^{(\mathbf{i})}-\sigma_{k}^{(\mathbf{i})}+m_{\text{adj}}^{(\mathbf{i})})^{2}}{(\sigma_{j}^{(\mathbf{i})}-\sigma_{k}^{(\mathbf{i})}-m_{\text{adj}}^{(\mathbf{i})})^{2}}
\prod_{l=1}^{N_{\mathbf{i}+1}}\dfrac{(\sigma_{l}^{(\mathbf{i})}-\sigma_{j}^{(\mathbf{i}+1)}-m_{\text{bfd}}^{(\mathbf{i,i+1})})^{2}}{(\sigma_{l}^{(\mathbf{i})}-\sigma_{j}^{(\mathbf{i}+1)}+m_{\text{bfd}}^{(\mathbf{i,i+1})})^{2}} \prod_{l=1}^{N_{\mathbf{i}-1}}\dfrac{(\sigma_{l}^{(\mathbf{i})}-\sigma_{j}^{(\mathbf{i}-1)}-m_{\text{bfd}}^{(\mathbf{i-1,i})})^{2}}{(\sigma_{l}^{(\mathbf{i})}-\sigma_{j}^{(\mathbf{i}-1)}+m_{\text{bfd}}^{(\mathbf{i-1,i})})^{2}}\\
&=\prod_{a=1}^{N_{f}^{(\mathbf{i})}}\dfrac{(\sigma_{j}^{(\mathbf{i})}+m_{a}^{(\mathbf{i})})^{2}}{(\sigma_{j}^{(\mathbf{i})}-m_{a}^{('\mathbf{i})})^{2}}
\end{aligned}
\end{equation}
Taking the square root of the vacuum equation at the $\mathbf{i}$-th node, we get the following vacuum equations with $N_{f}^{(\mathbf{i})}={N'}_{f}^{(\mathbf{i})}$ and $m_{a}^{(\mathbf{i})}={m'}_{a}^{(\mathbf{i})}$
\begin{equation}\label{40}
\begin{aligned}
&\prod_{k\neq j}^{N_{\mathbf{i}}}\dfrac{(\sigma_{j}^{(\mathbf{i})}-\sigma_{k}^{(\mathbf{i})}+m_{\text{adj}}^{(\mathbf{i})})}{(\sigma_{j}^{(\mathbf{i})}-\sigma_{k}^{(\mathbf{i})}-m_{\text{adj}}^{(\mathbf{i})})}
\prod_{l=1}^{N_{\mathbf{i}+1}}\dfrac{(\sigma_{l}^{(\mathbf{i})}-\sigma_{j}^{(\mathbf{i}+1)}-m_{\text{bfd}}^{(\mathbf{i,i+1})})}{(\sigma_{l}^{(\mathbf{i})}-\sigma_{j}^{(\mathbf{i}+1)}+m_{\text{bfd}}^{(\mathbf{i,i+1})})} \prod_{l=1}^{N_{\mathbf{i}-1}}\dfrac{(\sigma_{l}^{(\mathbf{i})}-\sigma_{j}^{(\mathbf{i}-1)}-m_{\text{bfd}}^{(\mathbf{i-1,i})})}{(\sigma_{l}^{(\mathbf{i})}-\sigma_{j}^{(\mathbf{i}-1)}+m_{\text{bfd}}^{(\mathbf{i-1,i})})}\\
&=\prod_{a=1}^{N_{f}^{(\mathbf{i})}}\dfrac{(\sigma_{j}^{(\mathbf{i})}+m_{a}^{(\mathbf{i})})}{(\sigma_{j}^{(\mathbf{i})}-m_{a}^{'(\mathbf{i})})}
\end{aligned}
\end{equation}
and
\begin{equation}\label{}
\begin{aligned}
&\prod_{k\neq j}^{N_{\mathbf{i}}}\dfrac{(\sigma_{j}^{(\mathbf{i})}-\sigma_{k}^{(\mathbf{i})}+m_{\text{adj}}^{(\mathbf{i})})}{(\sigma_{j}^{(\mathbf{i})}-\sigma_{k}^{(\mathbf{i})}-m_{\text{adj}}^{(\mathbf{i})})}
\prod_{l=1}^{N_{\mathbf{i}+1}}\dfrac{(\sigma_{l}^{(\mathbf{i})}-\sigma_{j}^{(\mathbf{i}+1)}-m_{\text{bfd}}^{(\mathbf{i,i+1})})}{(\sigma_{l}^{(\mathbf{i})}-\sigma_{j}^{(\mathbf{i}+1)}+m_{\text{bfd}}^{(\mathbf{i,i+1})})} \prod_{l=1}^{N_{\mathbf{i}-1}}\dfrac{(\sigma_{l}^{(\mathbf{i})}-\sigma_{j}^{(\mathbf{i}-1)}-m_{\text{bfd}}^{(\mathbf{i-1,i})})}{(\sigma_{l}^{(\mathbf{i})}-\sigma_{j}^{(\mathbf{i}-1)}+m_{\text{bfd}}^{(\mathbf{i-1,i})})}\\
&=-\prod_{a=1}^{N_{f}^{(\mathbf{i})}}\dfrac{(\sigma_{j}^{(\mathbf{i})}+m_{a}^{(\mathbf{i})})}{(\sigma_{j}^{(\mathbf{i})}-m_{a}^{'(\mathbf{i})})}
\end{aligned}
\end{equation}
Then we can match the vacuum equation (\ref{40}) with the Bethe ansatz equation (\ref{9}) with the dictionary
\begin{equation}
    \begin{aligned}
    & i \longleftrightarrow m_{\text{adj}}^{(\mathbf{i})},\quad \dfrac{i}{2} \longleftrightarrow m_{\text{bfd}}^{(\mathbf{i-1,i})}=m_{\text{bfd}}^{(\mathbf{i,i+1})}\\
    & -\theta_{a}^{(\mathbf{i})}+is_{a}^{(\mathbf{i})}\longleftrightarrow m_{a}^{(\mathbf{i})},\quad \theta_{a}^{(\mathbf{i})}+is_{a}^{(\mathbf{i})}\longleftrightarrow m_{a}^{'(\mathbf{i})}
\end{aligned}
\end{equation}

The vacuum equation of $A_{r}$ quiver gauge theory with $\text{SO}$ and $\text{Sp}$ gauge group is expected to correspond to Bethe ansatz equation of $sl_{r+1}$ open spin chain with the diagonal type boundary condition. For an $\text{Sp}(N_{\alpha})$ gauge group at the $\alpha$-th node, one vacuum equation can be written as
\begin{equation*}\label{}
\begin{aligned}
&\dfrac{\text{sin}^{2}(2\sigma_{j}^{(\alpha)}-m_{\text{adj}}^{(\alpha)})}{\text{sin}^{2}(2\sigma_{j}^{(\alpha}+m_{\text{adj}}^{(\alpha)})}\prod_{j\neq k}^{N_{\alpha}}\dfrac{\text{sin}(\sigma_{j}^{(\alpha)}\pm \sigma_{k}^{(\alpha)}-m_{\text{adj}}^{(\alpha)})}{\text{sin}(-\sigma_{j}^{(\alpha)}\pm \sigma_{k}^{(\alpha)}-m_{\text{adj}}^{(\alpha)})}\prod_{a=1}^{N_{f}^{(\alpha)}}\dfrac{\text{sin}^{2}(\sigma_{j}^{(\alpha)}-m_{a}^{(\alpha)})}{\text{sin}^{2}(-\sigma_{j}^{(\alpha)}-m_{a}^{(\alpha)})}\\
\end{aligned}
\end{equation*}
\begin{equation}
\begin{aligned}
&\times \prod_{j\neq k}^{N_{\alpha}}\dfrac{\text{sin}(\sigma_{j}^{(\alpha)}\pm \sigma_{k}^{(\alpha)}-m_{\text{adj}}^{(\alpha)})}{\text{sin}(-\sigma_{j}^{(\alpha)}\pm \sigma_{k}^{(\alpha)}-m_{\text{adj}}^{(\alpha)})}\prod_{a=1}^{{N'}_{f}^{(\alpha)}}\dfrac{\text{sin}^{2}(\sigma_{j}^{(\alpha)}-m_{a}^{(\alpha)})}{\text{sin}^{2}(-\sigma_{j}^{(\alpha)}-m_{a}^{(\alpha)})}\\
&\times\prod_{k=1}^{N_{\alpha-1}}\dfrac{\text{sin}^{2}(\sigma_{j}^{(\alpha)}\pm \sigma_{k}^{(\alpha-1)}-m_{\text{bfd}}^{(\alpha-1,\alpha)})}{\text{sin}^{2}(-\sigma_{j}^{(\alpha)}\pm \sigma_{k}^{(\alpha-1)}-m_{\text{bfd}}^{(\alpha-1,\alpha)})}\prod_{l=1}^{N_{\alpha+1}}\dfrac{\text{sin}^{2}(\sigma_{j}^{(\alpha)}\pm \sigma_{l}^{(\alpha+1)}-m_{\text{bfd}}^{(\alpha,\alpha+1)})}{\text{sin}^{2}(-\sigma_{j}^{(\alpha)}\pm\sigma_{l}^{(\alpha+1)}-m_{\text{bfd}}^{(\alpha,\alpha+1)})}=1
\end{aligned}    
\end{equation}
where we set the gauge group at the $(\alpha \pm 1)$-th node to be $\text{Sp}(N_{\alpha\pm 1})$; the other vacuum equation can be written 
\begin{equation}\label{}
\begin{aligned}
&\dfrac{\text{sin}^{2}(2\sigma_{j}^{(\alpha)}-m_{\text{adj}}^{(\alpha)})}{\text{sin}^{2}(2\sigma_{j}^{(\alpha}+m_{\text{adj}}^{(\alpha)})}\prod_{j\neq k}^{N_{\alpha}}\dfrac{\text{sin}(\sigma_{j}^{(\alpha)}\pm \sigma_{k}^{(\alpha)}-m_{\text{adj}}^{(\alpha)})}{\text{sin}(-\sigma_{j}^{(\alpha)}\pm \sigma_{k}^{(\alpha)}-m_{\text{adj}}^{(\alpha)})}\prod_{a=1}^{N_{f}^{(\alpha)}}\dfrac{\text{sin}^{2}(\sigma_{j}^{(\alpha)}-m_{a}^{(\alpha)})}{\text{sin}^{2}(-\sigma_{j}^{(\alpha)}-m_{a}^{(\alpha)})}\\
&\times \prod_{j\neq k}^{N_{\alpha}}\dfrac{\text{sin}(\sigma_{j}^{(\alpha)}\pm \sigma_{k}^{(\alpha)}-m_{\text{adj}}^{(\alpha)})}{\text{sin}(-\sigma_{j}^{(\alpha)}\pm \sigma_{k}^{(\alpha)}-m_{\text{adj}}^{(\alpha)})}\prod_{a=1}^{{N'}_{f}^{(\alpha)}}\dfrac{\text{sin}^{2}(\sigma_{j}^{(\alpha)}-m_{a}^{(\alpha)})}{\text{sin}^{2}(-\sigma_{j}^{(\alpha)}-m_{a}^{(\alpha)})}\\
&\times\prod_{k=1}^{N_{\alpha-1}}\dfrac{\text{sin}^{2}(\sigma_{j}^{(\alpha)}\pm \sigma_{k}^{(\alpha-1)}-m_{\text{bfd}}^{(\alpha-1,\alpha)})}{\text{sin}^{2}(-\sigma_{j}^{(\alpha)}\pm \sigma_{k}^{(\alpha-1)}-m_{\text{bfd}}^{(\alpha-1,\alpha)})}\prod_{l=1}^{N_{\alpha+1}}\dfrac{\text{sin}^{2}(\sigma_{j}^{(\alpha)}\pm \sigma_{l}^{(\alpha+1)}-m_{\text{bfd}}^{(\alpha,\alpha+1)})}{\text{sin}^{2}(-\sigma_{j}^{(\alpha)}\pm\sigma_{l}^{(\alpha+1)}-m_{\text{bfd}}^{(\alpha,\alpha+1)})}\\
&\times \dfrac{\text{sin}^{2\delta_{\alpha-1}}(\sigma_{j}^{(\alpha)}-m_{\text{bfd}}^{(\alpha-1,\alpha)})}{\text{sin}^{2\delta_{\alpha-1}}(-\sigma_{j}^{(\alpha)}-m_{\text{bfd}}^{(\alpha-1,\alpha)})}\dfrac{\text{sin}^{2\delta_{\alpha+1}}(\sigma_{j}^{(\alpha)}-m_{\text{bfd}}^{(\alpha,\alpha+1)})}{\text{sin}^{2\delta_{\alpha+1}}(-\sigma_{j}^{(\alpha)}-m_{\text{bfd}}^{(\alpha,\alpha+1)})}=1
\end{aligned}    
\end{equation}
where we set the gauge group at the $(\alpha \pm 1)$-th node to be $\text{SO}(2N_{\alpha\pm 1}+\delta_{\alpha\pm 1})$, $\delta_{\alpha\pm 1}$=0 or 1. For a node with $\text{SO}(2N_{\alpha}+\delta_{\beta})$ gauge group at the $\beta$-th side, we write the vacuum equation as
\begin{equation}\label{}
\begin{aligned}
&\dfrac{\text{sin}^{4\delta_{\beta}}(\sigma_{j}^{(\beta)}-m_{\text{adj}}^{(\beta)})}{\text{sin}^{4\delta_{\beta}}(\sigma_{j}^{(\beta}+m_{\text{adj}}^{(\beta)})}\prod_{j\neq k}^{N_{\beta}}\dfrac{\text{sin}(\sigma_{j}^{(\beta)}\pm \sigma_{k}^{(\beta)}-m_{\text{adj}}^{(\beta)})}{\text{sin}(-\sigma_{j}^{(\beta)}\pm \sigma_{k}^{(\beta)}-m_{\text{adj}}^{(\beta)})}\prod_{a=1}^{N_{f}^{(\beta)}}\dfrac{\text{sin}^{2}(\sigma_{j}^{(\beta)}-m_{a}^{(\beta)})}{\text{sin}^{2}(-\sigma_{j}^{(\beta)}-m_{a}^{(\beta)})}\\
&\times \prod_{j\neq k}^{N_{\beta}}\dfrac{\text{sin}(\sigma_{j}^{(\beta)}\pm \sigma_{k}^{(\beta)}-m_{\text{adj}}^{(\beta)})}{\text{sin}(-\sigma_{j}^{(\beta)}\pm \sigma_{k}^{(\beta)}-m_{\text{adj}}^{(\beta)})}\prod_{a=1}^{{N'}_{f}^{(\beta)}}\dfrac{\text{sin}^{2}(\sigma_{j}^{(\beta)}-m_{a}^{(\beta)})}{\text{sin}^{2}(-\sigma_{j}^{(\beta)}-m_{a}^{(\beta)})}\\
&\times\prod_{k=1}^{N_{\beta-1}}\dfrac{\text{sin}^{2}(\sigma_{j}^{(\beta)}\pm \sigma_{k}^{(\beta-1)}-m_{\text{bfd}}^{(\beta-1,\beta)})}{\text{sin}^{2}(-\sigma_{j}^{(\beta)}\pm \sigma_{k}^{(\beta-1)}-m_{\text{bfd}}^{(\beta-1,\alpha)})}\prod_{l=1}^{N_{\beta+1}}\dfrac{\text{sin}^{2}(\sigma_{j}^{(\beta)}\pm \sigma_{l}^{(\beta+1)}-m_{\text{bfd}}^{(\beta,\beta+1)})}{\text{sin}^{2}(-\sigma_{j}^{(\beta)}\pm\sigma_{l}^{(\beta+1)}-m_{\text{bfd}}^{(\beta,\beta+1)})}=1
\end{aligned}    
\end{equation}
We leave the comparison with Bethe ansatz equations for later work.

\section{Conclusions and discussion}\label{e}

In this article, we use the same approach in \cite{DZ23} to calculate the effective potential and the vacuum equations of $A_{2}$ quiver gauge theory with $\text{BCD}$-type product gauge group. Then we generalized the Bethe/Gauge correspondence proposed in \cite{DZ23} for the $\text{BCD}$-type gauge theories to $A_{2}$ quiver gauge theory with $\text{BCD}$-type product gauge group. And the corresponding open $\text{XXZ}$ spin chain with diagonal boundary condition on the Bethe side is modified to open $sl_{3}$ $\text{XXZ}$ spin chain with diagonal boundary condition. We saw that the correspondence worked perfect for 3d and 2d gauge theories with the parameter being specified to either $\xi=0,\frac{1}{2}$ or $\pm\frac{\eta}{2}$ to realize the vacuum equation of quiver gauge theory from the Bethe ansatz equation. Specially, the correspondence between 2d $A_{2}$ quiver gauge theory and $sl_{3}$ spin chain in \cite{KZ21} is one of our outcomes here. We also calculate the vacuum equation of general $A_{r}$ quiver gauge theory.  

For $A_{n}$ quiver gauge theory, we can change the classical Lie algebra into the exceptional Lie algebra to get the corresponding effective superpotential and vacuum equations. The Recently, the quantum integrability of non-simple laced quiver gauge theory has been discussed in \cite{CK18}. Meanwhile, the quantum integrable model of non-simple laced Lie algebra called folded integrable model has been given \cite{FHR22}. As for the integrability of 4d and 5d non-simple laced quiver gauge theory with $\text{SO}$ and $\text{Sp}$ gauge groups, as well as the $\text{SO/Sp}$ duality \cite{HKS11} seem to require more effort to study in the future. For the space of the paper, here we do not discus the Bethe/Gauge correspondence for quiver theories with gauge groups $\text{SU/SO}$ and $\text{SU/Sp}$. For that cases, the representations and the effective superpotential given in here and \cite{DZ23} are inevitably, we leave it to another work.


\acknowledgments
The financial support from the Natural Science Foundation of China (NSFC, Grants 11775299) is gratefully acknowledged from one of the authors (Ding).



\end{document}